\documentclass{article}

\usepackage{PRIMEarxiv}

\usepackage[utf8]{inputenc} 
\usepackage[T1]{fontenc}    
\usepackage{hyperref}       
\usepackage{url}            
\usepackage{booktabs}       
\usepackage{amsfonts}       
\usepackage{nicefrac}       
\usepackage{microtype}      
\usepackage{lipsum}
\usepackage{fancyhdr}       
\usepackage{graphicx}       
\graphicspath{{media/}}     

\usepackage{amsmath,amsfonts}
\usepackage{algorithmic}
\usepackage{algorithm}
\usepackage{array}
\usepackage[caption=false,font=normalsize,labelfont=sf,textfont=sf]{subfig}
\usepackage{textcomp}
\usepackage{stfloats}
\usepackage{url}
\usepackage{verbatim}
\usepackage{graphicx}
\usepackage{cite}
\usepackage{amssymb}
\usepackage{bm}
\usepackage{bbm}
\usepackage{color}
\usepackage{multirow}
\usepackage{diagbox}
\usepackage{threeparttable}

\newtheorem{remark}{Remark}[section]

\newtheorem{thm}{Theorem}[section]

\newtheorem{lemma}{Lemma}[section]
\newtheorem{proposition}{Proposition}[section]

\title{A ``Breathing'' Mobile Communication Network  
\thanks{This material is accepted as a regular paper in an upcoming issue of the IEEE Transactions on Mobile Computing, DOI: 10.1109/TMC.2024.3487213.
This material is
    supported by the Strategic Priority Research Program of Chinese Academy of Sciences under Grant No. XDA27030201, by the major key project of Peng Cheng Laboratory under grant PCL2023AS1-2, and by the National Natural Science Foundation of China under grants 12071465 and 11688101.} 
}

\author{
  Chao Ge \\
  Department of Electronic Engineering \\
  Tsinghua University \\
  Beijing\\
  \texttt{gechao@amss.ac.cn} \\
   \And
  Ge~Chen \\
  Key Laboratory of Systems and Control  \\
  Academy of Mathematics and Systems Science, Chinese Academy of Sciences \\
  Beijing\\
  \texttt{chenge@amss.ac.cn} \\
  \And
  Zhipeng Jiang \\
  School of Mathematical Sciences  \\
  University of Chinese Academy of Sciences \\
  Beijing\\
  \texttt{jiangzhipeng@ucas.ac.cn} 
}

\begin{document}
\maketitle

\begin{abstract}
The frequent migration of large-scale users leads to the load imbalance of mobile communication networks, which causes resource waste and decreases user experience.
To address the load balancing problem, this paper proposes a dynamic optimization framework for mobile communication networks inspired by the average consensus in multi-agent systems.
In this framework, all antennas cooperatively optimize their CPICH (Common Pilot Channel) transmit power in real-time to balance their busy-degrees. Then,
the coverage area of each antenna would change accordingly, and we call this framework a  ``breathing'' mobile communication network.
To solve this optimization problem, two algorithms named BDBA (Busy-degree Dynamic Balancing Algorithm) and BFDBA (Busy-degree Fast Dynamic Balancing Algorithm) are proposed.
Moreover, a fast network coverage calculation method is introduced, by which each antenna's minimum CPICH transmit power is determined under the premise of meeting the network coverage requirements.
Besides, we present the theoretical analysis of the two proposed algorithms' performance, which prove that all antennas' busy-degrees will reach consensus under certain assumptions.
Furthermore, simulations carried out on three large datasets demonstrate that our cooperative optimization can significantly reduce the unbalance among antennas as well as the proportion of over-busy antennas.
\end{abstract}


\section{Introduction}\label{intro}
\begin{figure}[htb]
    \centering
        \subfloat[]{
            \includegraphics[width=3.7in]{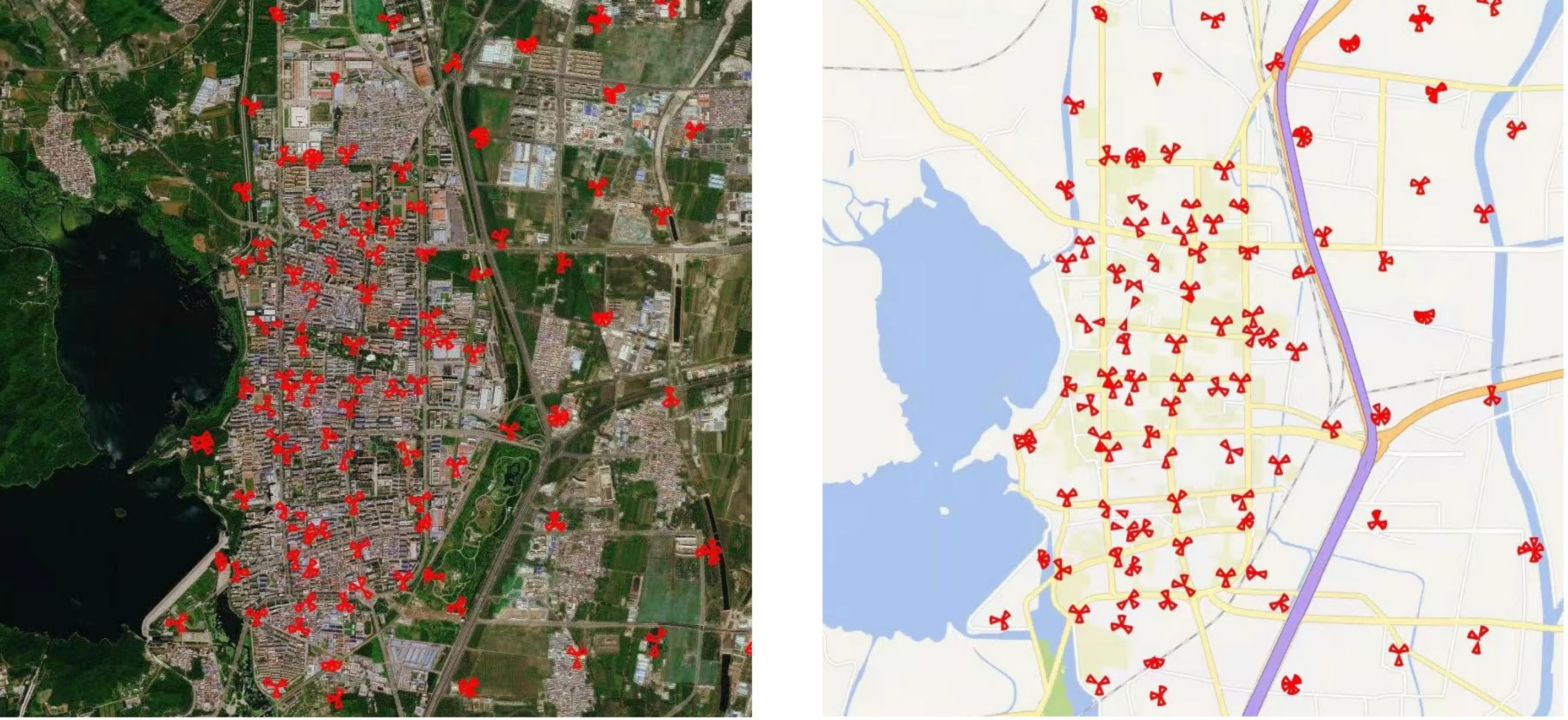}}
        \\
        \subfloat[]{
            \includegraphics[width=3.7in]{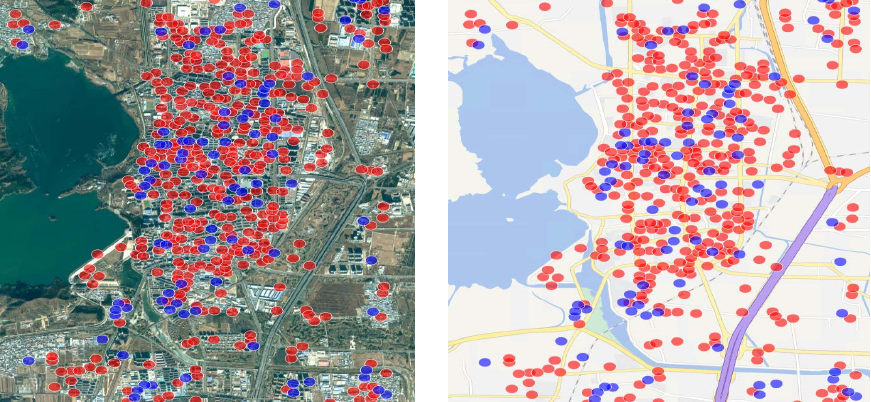}}
\caption{(a) The satellite map (left) and GPS map (right) of a selected area in Beijing, where the red icons denote the antennas; (b) the comparison of network traffic hotspots at different times on the satellite map (left) and the GPS map (right), where the red circles represent the hotspots during 10:00-11:00 period and the blue circles correspond to 22:00-23:00 period. The map data is provided by National Geographic Information Public Service Platform of China, and the antenna data is provided by Beijing Mobile Company.}
\label{Fig:map}
\end{figure}

Over the recent decades,
emerging communication technologies generate a significant increase in communication traffic as well as the use of related devices such as cellular BSs.
An accompanying challenge consists of addressing the ``tidal effect" of communication networks, which presents the huge fluctuation in traffic requirements caused by the frequent migration of large-scale users.
Especially in large cities, due to the dense and frequently migrating population, the tidal effect of mobile communication network traffic is more obvious \cite{pomp2016elastic,khan2015reducing,niu2011tango}.
People gather in commercial or business areas during working hours, and collectively migrate to residential areas after work hours which leads to the flow of network traffic.
To illustrate the ``tidal effect" phenomenon and its potential consequences, we consider Beijing as an example to intuitively show the real situation of mobile communication network.
We select a general area in Beijing, including both commercial and residential cells, which is shown in  Fig.\ref{Fig:map} (a).
Moreover, Fig.\ref{Fig:map} (b) presents the comparison of network traffic hotspots at different times. It can be observed that there is little overlap between hotspots between these two time periods.
Due to the large variation of network traffic, the utilizations of antenna resources also have huge spatiotemporal fluctuations and are seriously unbalanced.

\begin{figure}[htbp]
  \centering
    \subfloat[]{
      \includegraphics[height=2in]{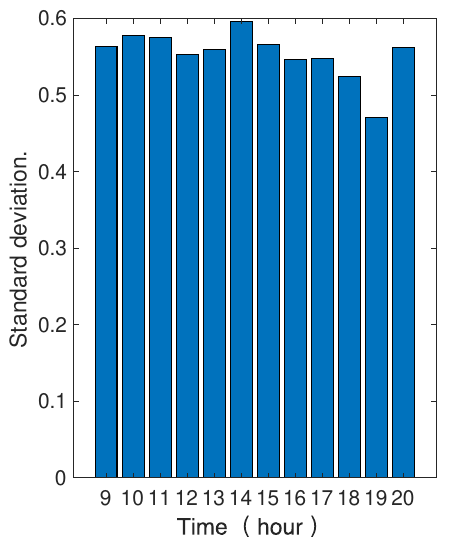}}
    \subfloat[]{
      \includegraphics[height=2in]{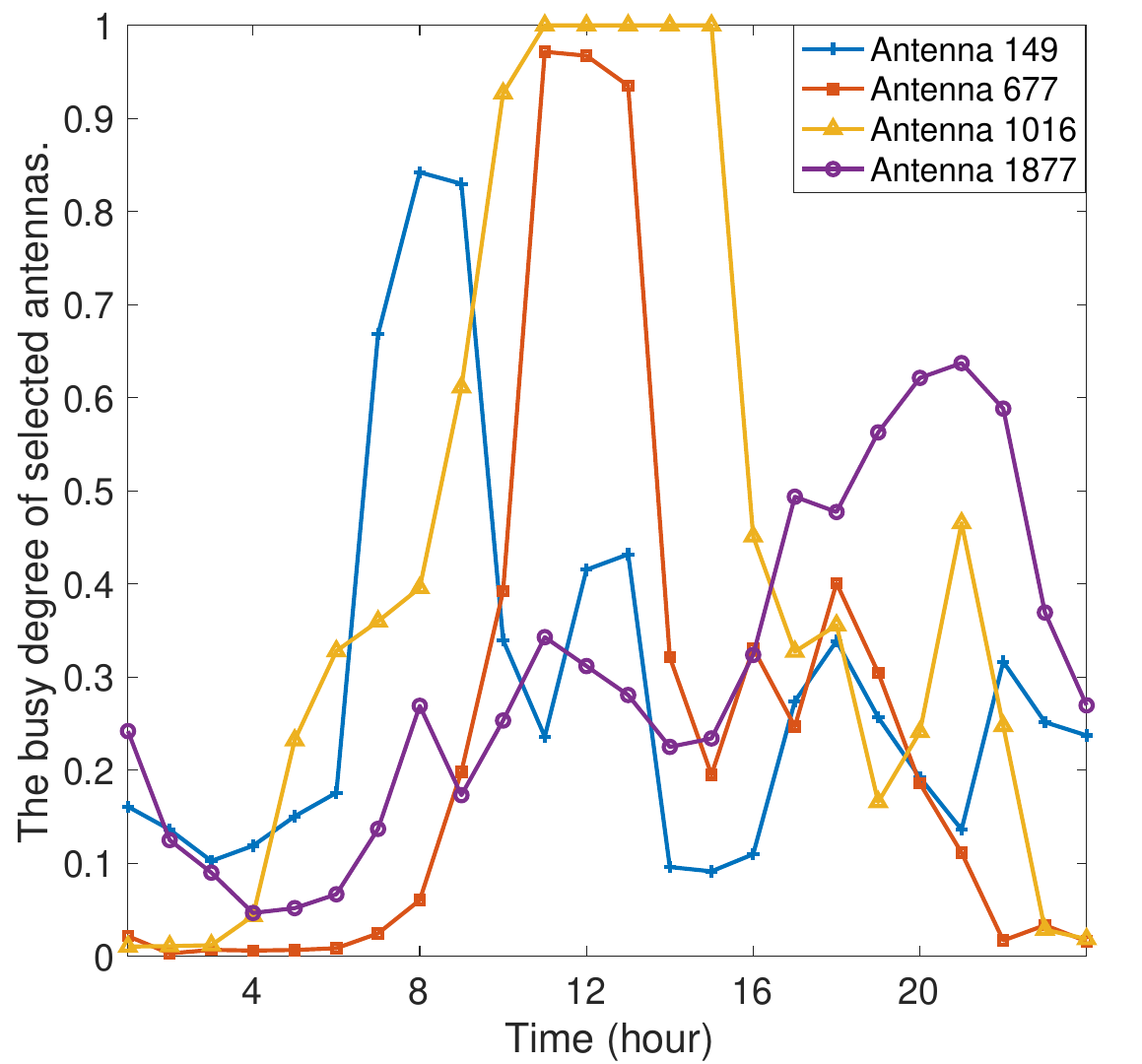}}
\caption{(a) The standard deviations of busy-degrees of all antennas in Fig.\ref{Fig:map} from 9:00 to 20:00. (b) The daily changes of randomly selected four antennas' busy-degrees. The MR and antenna data is provided by the Beijing Mobile Company, and the detailed calculation process is introduced in  Section \ref{simulation}.}
\label{Fig:antenna0}
\end{figure}

According to \cite{Xu2019,Nagib2021,Murudkar2019}, the PRB (physical resource block) utilization data can reflect the dynamic traffic load and resource utilization status in real mobile communication networks.
Hence, we measure the \emph{busy-degree} of each antenna by its average utilization rate of PRBs at different periods.
To more vividly illustrate the imbalance and huge fluctuations in the busy-degrees among antennas, we calculate the busy-degrees of all antennas in Fig.\ref{Fig:map} with the MR (measurement report) data from Beijing Mobile Company.
The detailed data information and calculation process are introduced in Section \ref{simulation}.
Fig. \ref{Fig:antenna0} (a) shows the standard deviations of the busy-degrees of all antennas, which illustrates the imbalance of antennas' busy-degrees.
Also, we randomly select four antennas in this area to observe the daily changes in their busy-degrees,
which are shown in Fig. \ref{Fig:antenna0} (b).
It can be observed that the busy-degrees of these antennas have large fluctuations over time, and there are also large differences among them at the same time.
However, the PRB resources are deployed according to the peak user demand, most PRBs are not efficiently utilized during most of the time, which will cause resource wastes and reduce the efficiency of the network \cite{khan2015reducing,niu2011tango,Mishra2014,huang2019geometric}.
Besides, the peak user demand affected by emergencies also has large fluctuation, which may cause some antennas are very busy or even congested at certain times, and the
user experience is reduced greatly.
Currently,
the tidal effect and the issues it engenders are major concerns of network operators \cite{auer2011much,cho2013energy}.
In particular, toward the upcoming 6G, the research scope needs to be extended to more complicated scenarios of integrating with satellite systems \cite{it2020toward}, dynamic optimization algorithms are in urgent need.


\subsection{Review of Related Literature}


Since the BS energy consumption accounts for most of the total network energy consumption, BS sleeping strategies have been widely studied, in which a BS should be turned off when the traffic it handles is low and the surrounding BSs
have enough resources to serve the users \cite{liu2015small,Guo2016delay,5992823,5683654,dawoud2014optimizing}.
Liu \emph{et al.} further design four different BS sleep modes and propose the optimization strategies \cite{liu2015small}.
Guo \emph{et al.} develop delay-constrained BS sleeping policies to optimize BS sleeping time and wake-up period \cite{Guo2016delay}.
These BS sleeping strategies can improve the network balance by sleeping idle antennas, however cannot solve the overload problem of busy antennas.

To address the overload problem caused by the  traffic imbalance, various research focuses on self-tuning  handover parameters, such as the \emph{cell individual offset} (CIO) \cite{Sheng2014,xu2019load,Attiah2020,Alsuhli2021,Alsuhli2021opti,Hasan2018,Alsuhli2023}.
Sheng \emph{et al.} propose a game-theoretic solution to solve the asymmetry traffic distribution among multiple cells \cite{Sheng2014}.
The deep reinforcement learning framework is exploited to optimize CIO values to better balance the traffic between different cells \cite{xu2019load,Attiah2020,Alsuhli2021,Hasan2018,Alsuhli2023}.
Further, Alsuhli \emph{et al.} jointly optimize the
transmission power and CIO values using reinforcement learning \cite{Alsuhli2021opti}.


Besides, the daily network traffic distribution during different periods is essential for load-aware dynamic network operations.
Hence, another major research focus is to analyze and predict the temporal and spatial characteristics of network traffic.
Dawoud \emph{et al.} introduce a hybrid traffic prediction model, based on linear regression, which forecasts the workload of the BSs by utilizing historic traffic traces \cite{dawoud2014optimizing}.
Moreover, some machine learning techniques are applied to predict the traffic load pattern over time \cite{wang2017spatiotemporal,wang2018spatio}.
Xu \emph{et al.} propose a scalable Gaussian process framework to achieve traffic prediction in a cost-efficient manner \cite{Xu2019wireless}.

In fact, the load balancing in mobile communication networks  can be treated as an average-consensus problem in multi-agent systems.
The average-consensus implies that a group of distinct nodes reach consensus by averaging the state of their neighbours at each time step.
This protocol has been
extensively studied in computer science, such as the distributed
coordination of mobile autonomous agents \cite{GC-ZL-LG:14,GC:17b} and networked systems \cite{GL-CG:17,8447261},  as well as the load balancing in parallel computers \cite{CYBENKO1989279}.
Inspired by the research of average-consensus protocols,  we consider the antennas of BSs as optimization units and aim to reach the balance (or consensus) of antennas' busy-degrees, and thus settle the load balancing problem.

\subsection{Statement of Our Main Contributions}

First, 
a dynamic cooperative optimization framework for mobile communication networks is innovatively proposed, where the CPICH transmit power of antennas is adjusted in real time to achieve the balance of antennas' busy-degrees.
As the coverage area of each antenna changes with the variation of its CPICH transmit power \cite{wang2012on,ts1996}, this optimization framework is called a ``breathing'' mobile communication network.
Under this framework, two dynamic optimization algorithms, BDBA (Busy-degree Dynamic Balancing Algorithm) and BFDBA (Busy-degree Fast Dynamic Balancing Algorithm), are proposed.
Compared to traditional optimization algorithms mostly based on prediction, our algorithms use the dynamic cooperation of antennas to balance their busy-degrees
according to the actual network traffic, which brings better real-time performance.
Moreover, our algorithms avoid long-time traffic prediction, and then they can significantly reduce the impact of the prediction errors over the optimization results.


Second, we introduce a fast network coverage calculation and optimization method based on MR data and the multi-layer perceptron (MLP).
The traditional methods for coverage calculation in mobile communication networks  \cite{sarkar2003survey,hamim2014overview} are mainly based on various wireless propagation models, which suffer from the large computational complexity and low accuracy drawbacks.
As a result, the existing optimization algorithms have difficulties in calculating the network coverage in real-time, thus the QoS is hard to guarantee.
Since MR data records the historical states of the users in the whole network, it can be used to train the MLP. With the well trained MLP, a fast computation can be provided for the coverage rate of the mobile communication network in real time.

Third, we give the theoretical analysis of the two proposed dynamic optimization algorithms' performance.
First, we study the property of the Jacobian matrix composed of the partial derivative of antennas' busy-degrees with respect to their CPICH transmit power and prove that it is a Laplacian matrix.
Using this property, we give the theoretical upper bound for the gap between BDBA's performance and global equilibrium.
Based on these findings, we specify the traffic distribution variation conditions for the busy-degrees of all antennas to reach global equilibrium (or consensus).
In addition, we propose a fast-solving algorithm BFDBA; estimate its theoretical upper bound for the gap, and show its consensus conditions.


Finally, we conducted extensive simulations in three general areas in Beijing with actual datasets from Beijing Mobile Company, which consist of one-day MR and antenna data.
The datasets correspond to the network state under static optimization.
The dataset-A includes $1956$ antennas and $112,629,717$ pieces of MR data.
Carrying out  BDBA and BFDBA on dataset-A,  the means of the standard deviations of all antennas' busy-degrees are reduced by $55.97\%$ and  $51.02\%$ respectively, provided that the coverage rate reaches $99.9\%$. Meanwhile, the mean of the proportions of over-busy antennas are declined by  $68.27\%$ and $65.03\%$ respectively under BDBA and BFDBA. Moreover, our algorithms have similar performance as dataset-A on dataset-B
and dataset-C, which contain $1658$ and $6120$ antennas, and $129,183,183$ and $266,485,384$ pieces of MR data respectively.

In order to highlight the significance of our work,  we give a comparison between our work and previous load balancing methods in  Table I.
From this table, it can be seen that our work has comprehensive advantages in terms of computational cost, theoretical analysis, interpretability, and scalability compared to previous studies.

\begin{table*}
\centering
\caption{Comparisons between our work and previous load balancing methods.}
\renewcommand\arraystretch{1.5}
\begin{threeparttable}
\begin{tabular}{|c|m{1.8cm}<{\centering}|m{1.6cm}<{\centering}|m{1.6cm}<{\centering}|m{1.2cm}<{\centering}|}
\hline
\diagbox{Algorithms and references}{Capabilities}   &  Computational Cost \tnote{1}   &  Theoretical  Analysis & Interpre -tability\tnote{2}  & Scalabi -lity \tnote{3}   \\
\hline
Zone-based mobility load balancing \cite{Sheng2014}   &  Uncertain & Yes & White-box & - \tnote{4} \\
\hline
DRL-based mobility load balancing\cite{xu2019load}   & High  & Yes &  Black-box & High\\
\hline
RL-based CIO optimization\cite{Attiah2020}    &  High  & No & Black-box & -\\
\hline
DRL-based CIO and energy control\cite{Alsuhli2021}   & High  & No & Black-box & - \\
\hline
RL-based CIO and power optimization\cite{Alsuhli2021opti}  & High  & No & Black-box & -  \\
\hline
Adaptive mobility load balancing\cite{Hasan2018}  & Low & No & White-box & -  \\
\hline
DRL-based CIO optimization\cite{Alsuhli2023}  & High  & No & Black-box & -  \\
\hline
Our work  & Low & Yes & White-box & High\\
\hline
\end{tabular}
    \begin{tablenotes}

       \item[1] We demonstrate the computational costs of algorithms by the time complexities.
      Given that the number of antennas is $n$, the algorithm in \cite{Sheng2014} needs to solve Nash equilibrium solutions of a non-cooperative and non-zero-sum game, which may have a complexity of PPAD (Polynomial Parity Argument on Directed graphs).
      The algorithm in  \cite{Hasan2018} primarily involves a series of computational and comparison operations based on current network conditions. Let $n_u$ be the number of users; the complexity is approximately $O(n n_u)$.
      As for the RL-based methods, the complexity generally depend on the size of the state space ($|\mathcal{S}|$), action space ($|\mathcal{A}|$), and the number of iterations required to converge ($T$).
      The time complexities are typically $O(|\mathcal{S}||\mathcal{A}|T)$, which often require many training episodes to achieve satisfactory performance.
      For the DRL-based  methods, the time complexity can become more complicated due to the training of neural networks.
      Hence, the methods in \cite{xu2019load,Attiah2020,Alsuhli2021,Alsuhli2021opti,Alsuhli2023} often come with high computational costs.
      As discussed in the following  Remark \ref{sample} and Subsection \ref{complexity_ana}, the complexity of our work is $O(n_s n_a n)$, where $n_s$ is the sampling volume and $n_a$ is all antennas' average number of neighbours.

      \item[2] Following \cite{Interpretability1,Interpretability2}, we measure the algorithms' scalability from the perspective of model transparency. According to the transparency of an algorithm’s internal logic or workings, the interpretability can be classified into white-box and black-box categories.  The game-theoretic framework \cite{Sheng2014}, event-driven adaptive control \cite{Hasan2018}, and our dynamic optimization framework are supported by well-defined models; therefore, these algorithms are classified as ``white-box". Different from model-based algorithms, machine learning methods \cite{xu2019load,Attiah2020,Alsuhli2021,Alsuhli2021opti,Alsuhli2023} remain opaque or hidden from comprehension; therefore, they are classified as ``black-box".
      \item[3] According to \cite{scalability1}, scalability is the ability of a system to handle an increasing number of elements, manage larger workloads smoothly, and adapt to expansion.  The authors in \cite{xu2019load}  emphasize the scalability of the proposed method due to the two-layer architecture. By performing random sampling on the MR data, our algorithms are also capable of rapid processing when scaled to large-scale scenarios, as demonstrated in the simulations.
      \item[4] ``-" denotes a lack of relevant analysis.
    \end{tablenotes}
  \end{threeparttable}
\end{table*}


\subsection{Paper Organization}\label{orga}
In Section \ref{description}, we present the preliminaries of our study.
We formulate our optimization problem and provide the solution scheme in Section \ref{problem}.
The detailed dynamic optimization algorithms, BDBA and BFDBA, and their theoretical analysis are proposed in Section \ref{algorithm} and Section \ref{convergence}.
The calculation method of the network coverage based on MR data and MLPs is then presented in Section \ref{coverage}.
In Section \ref{simulation}, we present simulations to test the capabilities of our algorithms. Finally, conclusion and future work are given in Section \ref{conclusion}.

\section{Preliminaries}\label{description}
To better model the optimization problem and introduce our algorithms, we present the preliminaries of our study which include the general mobile communication network model and introductions of busy-degree and network coverage rate in this section.
For convenience of understanding, the notations used in this paper are summarized in the following Table II.
\begin{table}
\centering
\caption{List of main notations.}
\renewcommand\arraystretch{1.5}
\begin{tabular}{m{3cm}<{\centering}|m{12cm}<{\centering}}
\hline
Notation   &  Description    \\
\hline
$\mathcal{N}$           &  Set of all antennas, $\mathcal{N}=\{1,\ldots,n\}$      \\
\hline
$\mathcal{N}_i$          & Set of neighbours of antenna $i$     \\
\hline
$\mathcal{M}_i$          & Set of MR data that antenna $i$ is the main service antenna\\
\hline
$p_i(\cdot)$         & CPICH transmit power of antenna $i$, $\vec{p}(\cdot)=\left(p_1(\cdot),\ldots,p_n(\cdot)\right)^{\top}$   \\
\hline
$u_i(\cdot)$         &  Theoretical adjustment of antenna $i$'s  CPICH transmit power, $\vec{u}(\cdot)=\left(u_1(\cdot),\ldots,u_n(\cdot)\right)^{\top}$   \\
\hline
$p_i^{\max}$         & Rated CPICH transmit power of antenna $i$   \\
\hline
$p_i^{\min}(\cdot)$         & Minimum CPICH transmit power of antenna $i$ which can satisfy the coverage requirement \\
\hline
$r_i$         &  Total of PRBs that allocated to antenna $i$   \\
\hline
$l_i(\cdot)$         & PRB utilization rate of antenna $i$   \\
\hline
$f_i(\cdot)$         & Busy-degree of antenna $i$   \\
\hline
$\bar{f}_i(\cdot)$         & Target busy-degree of antenna $i$   \\
\hline
$f_{i}^{\rm{rel}}(\cdot)$         & Relative busy-degree of antenna $i$, $\vec{f}^{\rm{rel}}(\cdot)=\left(f_1^{\rm{rel}}(\cdot),\ldots,f_n^{\rm{rel}}(\cdot)\right)^{\top}$   \\
\hline
$F(\cdot)$         &  Network coverage rate  \\
\hline
$F^{\rm{con}}$         &  Minimum requirement of network coverage rate  \\
\hline
$\mathcal{G} = (\mathcal{N},\mathcal{E}(k))$          &   Directed graph of all antennas with $\frac{1}{\bar{f_i}(k)}\frac{\partial f_i}{\partial p_j}(k)$
being the weight of edge $(i,j)$\\
\hline
$\mathcal{G}_f$        &    Graph of antennas failing to satisfy the coverage requirement\\
\hline
$A(\cdot)$          &   Jacobian matrix of linearization technique\\
\hline
$g_{\cdot}(\cdot): \mathbb{R}^n \mapsto \mathbb{R}$  &  Average traffic density function\\
\hline
$z(\cdot)$          &    Total of all antennas' average traffic\\
\hline
$\epsilon$          &   An adjustable constant used in $A(k)$'s approximation\\
\hline
$\gamma$          &   An adjustable constant controlling the magnitude of CPICH transmit power adjustment \\
\hline
$\tau$          &   An adjustable constant used in $\vec{u}^{(2)}(k)$'s calculation of FDOA\\
\hline
$\Delta p$          &   A minor adjustable constant\\
\hline
\end{tabular}
\end{table}

\subsection{General Mobile Communication Network Model}
This paper considers a general model of the mobile communication network. Let $\mathcal{R}$ be a servant region, which could be a city or a specific region. This region can be divided into several cells and each cell is corresponding to a BS antenna. This antenna is responsible for the mobile communication network control of its cell. In the classic MIMO BC (broadcast channel), each BS is equipped with multiple antennas\cite{weingarten2004capacity}.

In practical planning, instead of making major variations in network topology and layout, operators tend to change some configuration parameters of antennas to optimize the network usability.
The key adjustable parameters are at different difficulty levels, which can be divided into
the hard or soft network parameters.
The hard parameters mainly include network structure, base station deployment, and antenna's azimuth and tilt \cite{Chou2014,Smith1997,siomina2006automated}, which require some mechanical operation and cost highly in real-world scenarios.
In contrast, soft parameters are more suitable for highly frequent adjustment, mainly including CPICH transmit power, and CIO value \cite{Sheng2014,xu2019load,Attiah2020,Alsuhli2021,Alsuhli2021opti,Hasan2018,Alsuhli2023}.
Due to the easy transition from the CPICH transmit power adjustment to CIO adjustment, and the former is more easily extended to antenna sleep strategy by adjusting the CPICH transmission power of idle antennas to zero, we  employ the CPICH transmit power as the operation parameter to address the load balancing while ensuring network coverage.

We denote the set of all antennas in $\mathcal{R}$ by $\mathcal{N}=\{1,\ldots,n\}$, where $n$ is the total of antennas. Each antenna $i$ has a set of neighbours $\mathcal{N}_i$.
We consider that the neighbour relationship is symmetric, that is, if antenna $i$ is a neighbour of antenna $j$, then antenna $j$ must be a neighbour of antenna $i$. Cells are neighbours if their corresponding antennas are neighbours. Let $p_i(\cdot)$ denote antenna $i$'s CPICH transmit power. Following \cite{ahuja2014network}, the received signal strength is one of the most widely used parameters for network selection. It is assumed, through this paper, that each user accesses the antenna with the highest received CPICH signal strength. Hence, we can change the antennas' CPICH transmit power to adjust their CPICH signal strength received by the users. Then, the antennas accessed by the users are changed.
We set the sampling period to be $T$ and the time of each antenna is synchronized.
To better construct the model, we introduce the following indexes.

\subsection{Busy-degree}\label{des_busy}
To achieve the balance of the antennas' busy-degrees, we need to introduce an index that can describe the busy-degree of each antenna.
Inspired by \cite{Xu2019,jangsher2015joint}, the PRB utilization can reflect traffic flow changes and we can use the average PRB utilization rate to measure the busy-degree of each antenna in each sampling period.
To be more practical, we assume that the real-time statistics of the PRB utilization are measured in the timescale $H$ (typically in milliseconds).
Let $l_i(t)$ be the PRB number used by antenna $i$ at time $t$, and $\bar{l}_i[(k-1)T, kT)$ be the average number of PRBs used by antenna $i$ in the $k-$th sampling period $\big[(k-1)T, kT \big),~k\in \mathbb{Z}^+$,
i.e.,
\begin{equation}\label{networkload}
\bar{l}_i[(k-1)T, kT)=\frac{H}{T} \sum_{j=1}^{T/H} l_i \big( (k-1)T+(j-1)H \big).
\end{equation}
Let $r_i$ denote the total of PRBs that allocated to antenna $i$.
Then, the busy-degree of antenna $i$ in the $k-$th period $\big[(k-1)T, kT \big)$ is calculated by
\begin{equation}\label{networkload2}
f_i(k)=\bar{l}_i[(k-1)T, kT)/r_i,~~i\in \mathcal{N}.
\end{equation}
The calculation of each antenna's busy-degree can be carried out using clusters (such as an operator's cloud) in a centralized way or through the digital signal processors of each antenna in a distributed way.

Moreover, the CPICH transmit power of each antenna can affect its coverage area. The greater the CPICH transmit power is, the larger the  antenna's coverage area will be. Then, the busy-degree of each antenna can be regulated by adjusting its coverage area.
Therefore, we can consider $f_i(k)$ as a multivariate function of all antennas' CPICH transmit power, that is
\begin{equation}\label{f1}
f_i(k) = f_i\left(p_1(k),p_2(k),\dots,p_n(k),k\right),~i\in \mathcal{N}.
\end{equation}

\subsection{Network Coverage Rate}\label{NCrate}
Providing a high level of QoS to the users is the primary goal of mobile network operators. The network coverage rate is one of the main evaluation index of the QoS requirements\cite{semprebom2015sleep,kashi2012coverage}.
The received CPICH signal strength is a crucial factor to determine whether an area is covered or not.
To calculate the network coverage rate, various propagation models are proposed to calculate the path loss and the received signal strength \cite{sarkar2003survey,hamim2014overview}. However, the propagation of the wireless signals is affected by the reflection, the scattering, the diffraction, the transmission, as well as other factors, especially in the urban environment. Therefore, these propagation models are complex in calculation and low in accuracy; thus, they are not suitable for the real-time optimization algorithms of mobile communication networks.
To overcome these difficulties, we make use of the MR Data.
The MR data refers to the signal strength measurement report sent by users and reported to BSs.
Each piece of the MR data corresponds to a user in the network and contains several sets of received CPICH signal strength values and corresponding antenna IDs.
For example, $\vec{s}= \left( (I_1, s_1),(I_2, s_2),\ldots,( I_m, s_m) \right)$ is a piece of MR data sent by a user who received $m$ CPICH signal strength values,
where $s_1,\ldots, s_m$ are CPICH signal strength values and $I_1,\ldots,I_m$ are corresponding antenna IDs.
Obviously, $\left\{ I_1,I_2,\dots,I_m \right\} \subseteq \mathcal{N}$.
The MR data can monitor the whole network in real-time, allowing for more comprehensive and accurate evaluation of the mobile communication network \cite{shan2021intelligent}.

By \cite{wang2012on,ts1996}, we remark that the network coverage rate is positively correlated with antennas CPICH transmit power of antennas. We denote the network coverage rate of $\mathcal{R}$ in the $k$-th sampling period by $F\left(p_1(k),p_2(k),\ldots,p_n(k),k\right)$, and $F$ is a monotone increasing function of antennas' CPICH transmit power.
Based on this insight, we introduce a novel fast calculation and optimization method of network coverage rate with MR data and MLPs. The detailed algorithm is presented in Section \ref{coverage}.
The network coverage rate $F$ has a minimum requirement $F^{\rm{con}}$. If $F < F^{\rm{con}}$, the coverage requirement is not satisfied and we need to increase the CPICH transmit power of some antennas until $F \geqslant F^{\rm{con}}$.

\section{Problem Formulation}\label{problem}
Our goal is to dynamically adjust the CPICH transmit power of antennas, and then make their real-time busy-degrees reach their targets. At the same time, the network coverage rate needs to meet the requirement of QoS. In this section, we formally state the problem we are addressing in Subsection \ref{original}, then analyze the scheme of solving the problem in Subsection \ref{approximation} and Subsection \ref{solutionscheme}.

\subsection{Original Optimization Problem}\label{original}
Let $\bar{f_i}(k)$ denote the target busy-degree of antenna $i$ in the $k$-th sampling period and $u_1(k),\ldots,u_n(k)$ denote the adjustments of all antennas' CPICH transmit power in the $k$-th sampling period.
We aim to solve the $u_1(k),\ldots,u_n(k)$ that can make
\begin{equation}\label{dvec}
f_i\left(p_1(k)+u_1(k),\dots,p_n(k)+u_n(k),k\right) = \bar{f_i}(k),~i\in \mathcal{N}.
\end{equation}
Meanwhile, antennas' CPICH transmit power is limited by a
 maximal value, and the network coverage needs to be ensured.
We denote the maximal CPICH transmit power of antenna $i$ by $p_i^{\max}$.
Therefore, $u_1(k),\ldots,u_n(k)$ need to satisfy
\begin{equation}\label{lim}
\begin{cases}
p_i(k)+u_i(k) \leqslant p_i^{\max},~i\in \mathcal{N}\\
F\left(p_1(k)+u_1(k),\dots,p_n(k)+u_n(k),k\right) \geqslant F^{\rm{con}}
\end{cases}.
\end{equation}
Recall that $F^{\rm{con}}\in(0,1]$ is a constant denoting the minimum requirement of network coverage rate.
Our dynamic optimization problem is formulated by (\ref{dvec}) and (\ref{lim}).

\subsection{Approximation of Optimization Objective (\ref{dvec})}\label{approximation}
Since $f_i(k)$ is a complex nonlinear function without explicit expression that depends on the real-time network traffic and the CPICH transmit power $ p_1(k), \ldots, p_n(k)$,
it is very difficult to directly solve the equation (\ref{dvec}). We try to solve (\ref{dvec}) by linearization techniques.
By (\ref{f1}) and (\ref{dvec}), we have
$$
\bar{f_i}(k) - f_i(k) =  \frac{\partial f_i}{\partial p_1}(k) u_1(k)+\ldots+ \frac{\partial f_i}{\partial p_n}(k) u_n(k),~i\in \mathcal{N}.
$$
Since the network traffic volume remains unchanged, we only consider the case when
\begin{equation}\label{rank}
\sum_{i=1}^n\left(\bar{f_i}(k) - f_i(k)\right)=0,
\end{equation}
and the target busy-degree $\bar{f_i}(k)>0$ for any $i\in\mathcal{N},~k\in\mathbb{Z}^+$.
Then, we have
\begin{equation}\label{f2}
1-\frac{f_i(k)}{\bar{f_i}(k)} = \frac{1}{\bar{f_i}(k)}\left( \frac{\partial f_i}{\partial p_1}(k) u_1(k)+\ldots+ \frac{\partial f_i}{\partial p_n}(k) u_n(k)\right).
\end{equation}
Let
\begin{eqnarray*}
\vec{u}(k):=\left[\begin{array}{c}
u_1(k) \\
u_2(k) \\
\vdots \\
u_n(k)
\end{array}\right],~\vec{d}(k):=\left[\begin{array}{c}
1-\frac{f_1(k)}{\bar{f_1}(k)} \\
\vdots \\
1-\frac{f_n(k)}{\bar{f_n}(k)}
\end{array}\right],
\end{eqnarray*}
and
\begin{equation}\label{Amatrix}
A(k):=
\left[\begin{array}{cccc}
\frac{1}{\bar{f_1}(k)}\frac{\partial f_1}{\partial p_1}(k) & \frac{1}{\bar{f_1}(k)}\frac{\partial f_1}{\partial p_2}(k) & \cdots & \frac{1}{\bar{f_1}(k)}\frac{\partial f_1}{\partial p_n}(k) \\
\frac{1}{\bar{f_2}(k)}\frac{\partial f_2}{\partial p_1}(k) & \frac{1}{\bar{f_2}(k)}\frac{\partial f_2}{\partial p_2}(k) & \cdots & \frac{1}{\bar{f_2}(k)}\frac{\partial f_2}{\partial p_n}(k) \\
\vdots & \vdots & \ddots & \vdots \\
\frac{1}{\bar{f_n}(k)}\frac{\partial f_n}{\partial p_1}(k) & \frac{1}{\bar{f_n}(k)}\frac{\partial f_n}{\partial p_2}(k) & \cdots & \frac{1}{\bar{f_n}(k)}\frac{\partial f_n}{\partial p_n}(k)
\end{array}\right]
\end{equation}
be the Jacobian matrix.
By (\ref{f2}) and (\ref{Amatrix}), it should be
\begin{equation}\label{equ}
A(k) \vec{u}(k) = \vec{d}(k).
\end{equation}

\subsection{One Approximate Solution of Optimization Objective (\ref{dvec})}\label{solutionscheme}

Let $\mathcal{G}:= \mathcal{G}(k)= (\mathcal{N},\mathcal{E}(k))$ be the directed graph of all antennas, where $\mathcal{E}(k)$ is the edge set. For $i,j \in \mathcal{N}$, if $|\frac{\partial f_i}{\partial p_j}(k)|>0$,
the ordered pair $(i,j)$ is in the $\mathcal{E}(k)$, which implies that antenna $j$'s CPICH transmit power can affect antenna $i$'s busy-degree.
Assume that $\mathcal{G}$ is strongly connected,
according to the following Lemma \ref{lem} in Section~\ref{convergence}, the Jacobian matrix $A(k)$ is a Laplacian matrix with rank $n-1$.
Hence, $\mathrm{rank}(A(k),\vec{d}(k)) \geq n-1$.
On the other hand,
by (\ref{rank}), we have for any $i\in \mathcal{N}$,
$$\sum_{j=1}^n \frac{\partial f_{j}}{\partial p_i}(\tilde{p},k) =\frac{\partial \sum_{j=1}^n f_{j}}{\partial p_i}(\tilde{p},k)=-\frac{\partial\sum_{j=1}^n \bar{f}_{j}}{\partial p_i}(\tilde{p},k)=0.$$
We note that
\begin{align*}
\mathrm{rank}(A(k),\vec{d}(k))
&=\mathrm{rank}\left[\begin{array}{cccc}
\frac{1}{\bar{f_1}(k)}\frac{\partial f_1}{\partial p_1}(k) & \cdots & \frac{1}{\bar{f_1}(k)}\frac{\partial f_1}{\partial p_n}(k) & 1-\frac{f_1(k)}{\bar{f_1}(k)}  \\
\vdots & \vdots & \vdots & \vdots \\
\frac{1}{\bar{f_n}(k)}\frac{\partial f_n}{\partial p_1}(k) & \cdots & \frac{1}{\bar{f_n}(k)}\frac{\partial f_n}{\partial p_n}(k) & 1-\frac{f_n(k)}{\bar{f_n}(k)}
\end{array}\right]\\
&=\mathrm{rank}\left[\begin{array}{cccc}
\frac{\partial f_1}{\partial p_1}(k) & \cdots & \frac{\partial f_1}{\partial p_n}(k) & \bar{f_1}(k)-f_1(k) \\
\vdots & \vdots & \vdots & \vdots \\
\frac{\partial f_n}{\partial p_1}(k) & \cdots & \frac{\partial f_n}{\partial p_n}(k) & \bar{f_n}(k)-f_n(k)
\end{array}\right].
\end{align*}
By applying elementary row operation of adding all rows to the last row, we have
\begin{equation*}
\mathrm{rank}(A(k),\vec{d}(k))
=\mathrm{rank}\left[\begin{array}{cccc}
\frac{\partial f_1}{\partial p_1}(k) & \cdots & \frac{\partial f_1}{\partial p_n}(k) & \bar{f_1}(k)-f_1(k) \\
\vdots & \vdots & \vdots & \vdots \\
0 & \cdots & 0 & 0
\end{array}\right]
\leq n-1.
\end{equation*}
Therefore, $\mathrm{rank}(A(k),\vec{d}(k))=n-1$.
According to the Section 2.2 of \cite{strang2006linear},
the equation (\ref{equ}) is solvable.
By Lemma 6.5 of \cite{bullo2020lectures}, the non-zero eigenvalues of $A(k)$ have strictly-positive real part.
Combining the Appendix C of \cite{strang2006linear},
$A(k)$ has the singular value decomposition form
$$A(k)=U(k)\Lambda (k)V(k)^{\top},$$
where
\begin{equation*}
\Lambda(k)=
\left[\begin{array}{cccc}
\sigma_1(k) & \cdots & 0 & 0 \\
\vdots & \ddots & \vdots & \vdots \\
0 & \cdots & \sigma_{n-1}(k) & 0 \\
0 & \cdots & 0 & 0
\end{array}\right]:=
\left[\begin{array}{cc}
\Lambda_1(k) & 0 \\
0  & 0
\end{array}\right],
\end{equation*}
and $\sigma_s(k)=\sqrt{\lambda_s(A(k)^{\top}A(k))}=\sqrt{\lambda_s(A(k)A(k)^{\top})}>0,~s=1\cdots,n-1,$ are ranged in descending order.
The orthogonal matrix $U(k)\in\mathbb{R}^{n\times n}$ has corresponding orthonormal eigenvectors of $A(k)A(k)^{\top}$ and the orthogonal matrix $V(k)\in\mathbb{R}^{n\times n}$ has corresponding orthonormal eigenvectors of $A(k)^{\top}A(k)$.
Let $U(k):=\left(U_1(k),U_2(k)\right)$ and $V(k):=\left(V_1(k),V_2(k)\right)$, where $U_1(k),V_1(k) \in \mathbb{R}^{n\times(n-1)}$.
We note that
\begin{equation}\label{null}
A(k)\bm{1}_n= \bm{0}_n,
\end{equation}
where $\bm{1}_n\in \mathbb{R}^n=(1,\cdots,1)^{\top}$ and $\bm{0}_n\in \mathbb{R}^n=(0,\cdots,0)^{\top}$.
Hence, the nullspace of $A(k)$ is $\{c\bm{1}_n,~c\in\mathbb{R}\}$ and
\begin{equation}\label{nullvec}
U_2(k)=V_2(k)=\frac{1}{n}\bm{1}_n.
\end{equation}
Then, we have $$A(k)=U_1(k)\Lambda_1(k)V_1(k)^{\top}.$$
By the Appendix C of \cite{strang2006linear}, the pseudoinverse of $A(k)$ can be denoted by
\begin{equation}\label{pseudo}
A^{+}(k)=V_1(k)\Lambda_1^{-1}(k)U_1(k)^{\top}.
\end{equation}
Then,
\begin{equation}\label{solution1}
\vec{u}^*(k)=A^{+}(k)\vec{d}(k)=V_1(k)\Lambda_1^{-1}(k)U_1(k)^{\top}\vec{d}(k)
\end{equation}
is a solution of (\ref{equ}).
By (\ref{null}), $\{\vec{u}^*(k)+c\bm{1}_n,~c\in\mathbb{R}\}$ are the solutions of (\ref{equ}).
To ensure the uniqueness of the solution and minimize adjustments as much as possible, we solve the following equation system
\begin{equation}\label{system}
\begin{cases}
A(k) \vec{u}(k) = \vec{d}(k)\\
\bm{1}_n^{\top}\vec{u}(k) =0
\end{cases}.
\end{equation}
By (\ref{nullvec}) and (\ref{solution1}), $\bm{1}_n^{\top}\vec{u}^*(k) = 0.$
Therefore, $\vec{u}^*(k)$ is the desired unique solution of (\ref{system}),
and also an approximate solution of the original optimization objective (\ref{dvec}) by the discussion in Subsection \ref{approximation}.

\section{Algorithms for Solving Optimization Problem}\label{algorithm}
This section provides the approximation algorithms to solve our optimization problem (\ref{dvec}) and (\ref{lim}).
We note that the solution of (\ref{dvec}) is the theoretical adjustment of antennas' CPICH transmit power which can settle the unbalanced busy-degrees among antennas. Therefore, we consider solving (\ref{dvec}) first and then correcting the theoretical adjustments of antennas' CPICH transmit power according to the constraints (\ref{lim}).

By Subsection \ref{solutionscheme}, if we get $A(k)$ then the approximate solution of optimization objective (\ref{dvec}) is obtained.
However,
as stated in Subsection \ref{approximation}, $f_i(k)$ is a complex nonlinear function and the partial derivatives $\frac{\partial f_i}{\partial p_j}(k),~i,j \in \mathcal{N}$, are difficult to solve directly.
To overcome this problem, we propose a novel method with MR data to get the approximations of the partial derivatives.
On this basis, we propose two methods to solve our optimization problem approximately by the formula (\ref{solution1}) and  the constraints (\ref{lim}).
Our two dynamic optimization algorithms are referred to as BDBA (Busy-degree Dynamic Balancing Algorithm) and BFDBA (Busy-degree Fast Dynamic Balancing Algorithm).
The flow chart of these two algorithms are presented in the following Fig. \ref{Fig:flow}, while the detailed algorithm processes are introduced in Subsections \ref{BDBA}
and \ref{BFDBA}.

\begin{figure}[htbp]
\centering
\includegraphics[width=4in]{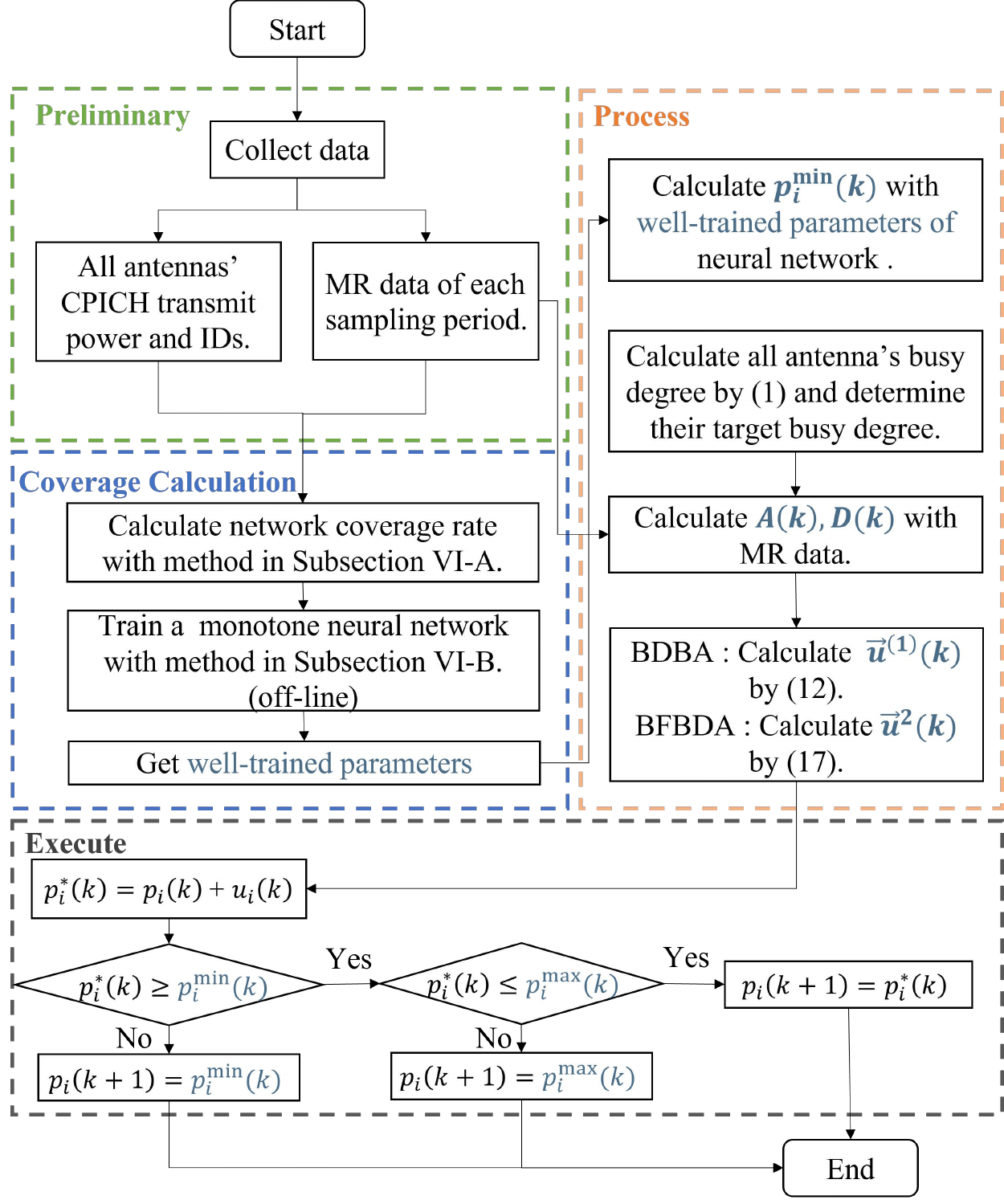}
\caption{The flow chart of BDBA and BFDBA.}
\label{Fig:flow}
\end{figure}

\subsection{Busy-degree Dynamic Balancing Algorithm}\label{BDBA}
First, we figure out the target busy-degree of each antenna. The target busy-degree $\bar{f_i}(k)$ has different calculation methods according to different algorithmic requirements.
For example, if we aim at global optimization, $\bar{f_i}(k)^{(1)}$ can be represented by
$$\bar{f_i}(k)^{(1)} = \frac{\sum_{j=1}^{n} r_j f_j(k)}{\sum_{j=1}^{n} r_j}.$$
If we consider local optimization, $\bar{f_i}(k)^{(2)}$ can be represented by
$$\bar{f_i}(k) ^{(2)}= \frac{r_i f_i(k)+\sum_{j\in N_i}r_j f_j(k)}{r_i + \sum_{j\in N_i} r_j}.$$
For each sampling period $k = 1, 2,\dots$, we perform the following steps in turn. \\
\textbf{Step 1 :} Calculate antenna $i$'s busy-degree $f_i(k)$ with (\ref{networkload2}).
\\
\textbf{Step 2 :} Approximate the partial derivative of $f_i$ with respect to $p_j$ with MR data of the $k$-th sampling period.

In practice, for $\vec{s} = \left( (I_1, s_1),(I_2, s_2),\ldots,( I_m, s_m) \right)$,  the first antenna  $I_1$ in $\vec{s}$ is its main service antenna.
First, we filter out the MR data in which its main service antenna's CPICH signal strength is lower than other antennas'.
Then, we select those pieces of MR data that antenna $i$ is the main service antenna and denote them by $\mathcal{M}_i(k)$.
For each piece of MR data in $\mathcal{M}_i(k)$, we note that the received CPICH signal strength corresponding to antenna $i$ is greater than or equal to other antennas'.
Let $\epsilon > 0$ be a constant.
After reducing antenna $i$'s CPICH signal strength by $\epsilon p_i(k)$,
we let $\delta^{-}_{i,j}(k)$ denote the number of MR data in which antenna $j$'s CPICH signal strength becomes the maximum.
Similarly, $\delta^{+}_{i,j}(k)$ denotes the number of MR data in which antenna $j$'s CPICH signal strength becomes the maximum after increasing antenna $j$'s CPICH signal strength by $\epsilon p_j(k)$. Then, we approximate $\partial f_i /\partial p_j(k)$ by
\begin{equation}\label{appro}
\begin{aligned}
\frac{\partial f_i}{\partial p_i}(k) & \approx \frac{\frac{f_i(k)}{\left|\mathcal{M}_i(k)\right|} \times \sum_{j \ne i} \delta_{i,j}^{-}(k)+\sum_{j\ne i} \frac{\delta_{j,i}^{+}(k) \times f_j(k) \times r_j}{\left|\mathcal{M}_j(k)\right| \times r_i}}{2 \epsilon  p_i(k)}  =:\bar{f}_i(k)\widetilde{A}_{ii}(k), \\
\frac{\partial f_i}{\partial p_j}(k) & \approx \frac{-\frac{\delta_{i, j}^{+}(k)}{\left|\mathcal{M}_i(k)\right|}-\frac{\delta_{j,i}^{-}(k) \times f_j(k) \times r_j}{\left|\mathcal{M}_j(k)\right| \times r_i}}{2 \epsilon  p_j(k)} =:\bar{f}_i(k)\widetilde{A}_{ij}(k),
\end{aligned}
\end{equation}
where $|\cdot|$ denotes the cardinality for a set. Therefore, we can obtain the approximated matrix
$\widetilde{A}(k):=[\widetilde{A}_{ij}(k)]\in\mathbb{R}^{n\times n}.$
\\
\textbf{Step 3 :} Calculate the theoretical adjustments of all antennas' CPICH transmit power.
With the approximation $\widetilde{A}(k)$, we can calculate $\widetilde{A}^{+}(k)$ by (\ref{pseudo}) and
let $\vec{u}^{(1)}(k) = \widetilde{A}^{+}(k)\vec{d}(k);$ or we can get the solution $\vec{u}^{(1)}(k)$ by solving (\ref{system}), i.e.,
\begin{equation}\label{system-add}
\begin{cases}
\widetilde{A}(k) \vec{u}^{(1)}(k) = \vec{d}(k)\\
\bm{1}_n^{\top}\vec{u}^{(1)}(k) =0
\end{cases}.
\end{equation}
%
\textbf{Step 4 :} Correct the theoretical adjustments of antennas' CPICH transmit power according to the constraints (\ref{lim}).\\
For antenna $i$, let
\begin{equation*}\label{gammaeq}
p^*_i(k):= p_i(k) + \gamma u_i^{(1)}(k),
\end{equation*}
where $0<\gamma\leq 1$ is a smoothing factor which controls the magnitude of the adjustment.
We denote the minimum CPICH transmit power of  antenna $i$ in the $k$-th sampling period which can satisfy the coverage requirement by $p_i^{\min}(k)$. The calculation of $p_i^{\min}(k)$ is then presented in Section \ref{coverage}. Then, the CPICH transmit power of antenna $i$ in the $(k+1)$-th sampling period is updated by
\begin{equation}\label{adjust}
p_i(k+1) =\left\{
\begin{aligned}
	&p_i^{\max}    ,~~~~~p^*_i(k)>p_i^{\max}\\
	&p^*_i(k)      ,~~~~p_i^{\min}(k) \leqslant p^*_i(k)\leqslant p_i^{\max} \\
	&p_i^{\min}(k),~~p^*_i(k) < p_i^{\min}(k)
\end{aligned}
\right..
\end{equation}

These steps can be carried out by clusters (such as an operator's cloud) in a centralized way or by the digital signal processors of each antenna in a distributed way.

\begin{remark}\label{sample}
Because the MR datasets $\mathcal{M}_i(k), i=1,\ldots,n,$ are usually very large,
the most computationally intensive part in Step 2 is the calculation of $\delta^{-}_{i,j}(k)/|\mathcal{M}_i(k)|$ and $\delta^{+}_{i,j}(k)/|\mathcal{M}_i(k)|$, whose computational complexities
are both $O(|\mathcal{M}_i(k)|)$. However, we can use the sampling method to approximate these calculations. In other words,
we can randomly choose $n_s$ pieces of MR data from each $\mathcal{M}_i(k)$ to approximatively calculate $\delta^{-}_{i,j}(k)/|\mathcal{M}_i(k)|$ and $\delta^{+}_{i,j}(k)/|\mathcal{M}_i(k)|$, whose computational complexities
are both $O(n_s)$.
It is valuable to remark that these computations are highly parallelizable.
\end{remark}

\begin{remark}
To avoid drastic  transitions of the access network that could affect user experience, we add the smoothing factor $\gamma$ in Step 4 to control the transition magnitude.
The smaller the value of $\gamma$, the smaller the transition magnitude, however the balancing effect of our algorithm will also deteriorate. The smoothing factor $\gamma$ can serve as a tunable parameter for characterizing this tradeoff.
\end{remark}

\subsection{Busy-degree Fast Dynamic Balancing Algorithm}\label{BFDBA}
We note that the approximation of $A(k)$ is time-consuming, therefore we present a fast dynamic optimization algorithm that only uses the approximation of $\frac{\partial f_i}{\partial p_i}(k)$. BFDBA makes a small change to the calculation methods for $\vec{u}(k)$ in \textbf{Step 3}.

Let
\begin{eqnarray*}
\vec{d}^{(2)}(k):=\left[\begin{array}{c}
1-\frac{f_1(k)}{\bar{f_1}(k)}-\tau p_1(k) \\
\vdots \\
1-\frac{f_n(k)}{\bar{f_n}(k)}-\tau p_n(k)
\end{array}\right],
\end{eqnarray*}
where $\tau>0$ is an adjustable parameter.
Let $D(k)$ be the diagonal matrix of $A(k)$, and choose
\begin{equation}\label{solu2}
\vec{u}^{(2)}(k) = D^{-1}(k)\vec{d}^{(2)}(k).
\end{equation}
Let $\widetilde{D}(k)$ be the diagonal matrix of $\widetilde{A}(k)$,
then
$$
\vec{u}^{(2)}(k) = \widetilde{D}^{-1}(k)\vec{d}^{(2)}(k).
$$

The remaining steps are the same as BDBA.

\subsection{Analysis of Computational Complexity for BDBA and BFDBA} \label{complexity_ana}

We first provide a brief analysis for the computational complexity of BDBA.
The computational complexity of Step 1 is very small and can be omitted. For Step 2,
we assume that the average number of neighbours for all antennas is $n_a$, and adopt the sampling strategy outlined in Remark \ref{sample}.
Then, the computational complexity of Step 2 is $O(n_s n_a n)$.
For Step 3, because the matrix $\widetilde{A}(k)$ is a sparse matrix, we can use the relaxation method to solve (\ref{system-add}), whose
computational complexity is $O(T_3 n_a n)$, where $T_3$ is the number of iterations.
For Step 4,
the calculation is dominated by determining $p_i^{\min}$, whose computational complexity is approximately equivalent to that of $F_i^{\star}$.
According to the following Remark \ref{rem-complex},
the computational complexity of $F_i^{\star}$ by the MLP is   $O(\sum_{l=0}^{m}N_{\ell}N_{\ell+1})$, where $N_{\ell}$ is the
number of neurons in hidden layer $\ell$, and $m$ is the number of hidden layers. Because the values of $T_3$, $N_{\ell}$ and $m$ are generally not big,
the dominant term of the overall computational complexity is $O(n_s n_a n)$.

As for BFDBA, the difference compared to BDBA lies in Step 3, where the complexity of inverting a diagonal matrix is $O(n)$.
Hence, the overall computational complexity of BFDBA remains $O(n_s n_a n)$.

\section{Theoretical Analysis}\label{convergence}
To prove the effectiveness of the proposed dynamic optimization algorithms, this section gives the theoretical analysis.
We first give the upper bound of the difference between the control effect of our algorithms and the global equilibrium state.
Also, we prove that under the control of our algorithms, all antennas' busy-degrees can achieve global equilibrium or approximately equilibrium based on some conditions concerning the traffic distribution variation.

A core issue of the analysis is the spatial distribution of network traffic, which has been modeled by various methods\cite{lee2013stochastic,guo2013spatial,riihijarvi2010modeling}. These models are generally based on some assumptions, such as the traffic distribution is  continuous  and follows some spatial stochastic distribution\cite{lee2013stochastic,guo2013spatial,riihijarvi2010modeling}.
Following previous works, we assume the average traffic distribution during each sampling period is continuous.
However, we don't have to assume any specific traffic distribution.

At any point $x$ on the map, the received CPICH transmit power of antenna $i$ equals antenna $i$'s CPICH  transmit power minus the attenuation value during propagation in decibel; therefore, we assume that the average traffic distribution is a function with respect to the attenuation vector of $x$, whose elements are the attenuation values of CPICH signals from all antennas to $x$.
Set $\vec{a}:=(a_1,a_2,\ldots,a_n)^\top \in \mathbb{R}^n$ to be the attenuation vector corresponding to all antennas.
As we mentioned in Section~\ref{description}, we assume that each user accesses the antenna corresponding to the highest received CPICH signal strength.
Let $g_k(\vec{a}): \mathbb{R}^n \mapsto \mathbb{R}$ denote the \emph{average traffic density function} regarding the point with attenuation vector $\vec{a}$ during the $k$-th sampling period $\left[(k-1)T,kT\right)$,
which satisfies
\begin{equation}\label{f_def2a}
\int_{\mathbb{R}^n} g_k(\vec{a}) \mathbbm{1}_{\left\{p_i(k)- a_i \geq p_j(k)-a_j,\forall j\right\}} \mathrm{d}\vec{a}
=\bar{l}_i[(k-1)T, kT).
\end{equation}
Following the assumption of the continuous distribution, we assume $g_k(\vec{a})$ is a continuous function with respect to $\vec{a}$.
Combining (\ref{f_def2a}) with (\ref{networkload2}) we have
\begin{equation}\label{f_def2}
f_{i}(k) = \frac{1}{r_i} \int_{\mathbb{R}^n} g_k(\vec{a}) \mathbbm{1}_{\left\{p_i(k)- a_i \geq p_j(k)-a_j,\forall j\right\}} \mathrm{d}\vec{a},
\end{equation}
where $\mathbbm{1}_{\{\cdot\}}$ is the characteristic function.

Let $z(k)$ be the total of all antennas' average traffic during the $k$-th sampling period.
Then, by the definition of $g_k(\vec{a})$, we have
\begin{equation*}
z(k) = \int_{\mathbb{R}^n} g_k(\vec{a}) \mathrm{d} \vec{a}.
\end{equation*}
We denote the weight of edge $(i,j)$ in graph $\mathcal{G}(k)$ by $\frac{1}{\bar{f_i}(k)}\frac{\partial f_i}{\partial p_j}(k)$.
First, we give the property of $A(k)$ in the following Lemma \ref{lem}. The proof is involved and is deferred to the Appendix in the supplementary file.

\begin{lemma}\label{lem}
Suppose that antenna $i$'s busy-degree takes the value by (\ref{f_def2}), and the average traffic density function $g_k(\vec{a})$ is continuous on $\mathbb{R}^n, ~k\in \mathbb{Z}^+$,
if the graph $\mathcal{G}$ is strongly connected and the elements of its out-degree matrix are positive,
then $A(k)$ defined in (\ref{Amatrix}) is a Laplacian matrix and $\mathrm{rank}(A(k))=n-1$.
\end{lemma}

Based on Lemma \ref{lem}, we give the theoretical analysis for the proposed BDBA and BFDBA.
By (\ref{adjust}), the CPICH transmit power of  antenna $i$ is limited by $p_i^{\max}$ and $p_i^{\min}(k)$.
To simplify the analysis, we assume that $p^*_i(k),~i \in \mathcal{N},$ is always not bigger than $p_i^{\max}$ and not smaller than $p_i^{\min}(k)$. Therefore, the equation (\ref{adjust}) can be rewritten as
\begin{equation}\label{equ3}
\vec{p}(k+1) = \vec{p}(k)+\gamma \vec{u}(k).
\end{equation}
Besides, we consider the global optimization and let the target busy-degree $$\bar{f}_i(k) = \bar{f_i}(k)^{(1)}= \frac{z(k)}{\sum_{j=1}^n r_j},$$
which is also the consensus state of all antennas' busy-degrees.
Then, by (\ref{dvec}) we have
\begin{align}\label{dnew}
&\vec{d}(k)=\vec{d}\left(\vec{p}(k),k\right)=\left[\begin{array}{c}
1 - \frac{f_{1}\left(\vec{p}(k),k\right)}{\frac{z(k)}{\sum_{j=1}^n r_j}} \\
\vdots \\
1 - \frac{f_{n}\left(\vec{p}(k),k\right)}{\frac{z(k)}{\sum_{j=1}^n r_j}}
\end{array}\right],\\\nonumber
\end{align}
and
\begin{equation}\label{anew}
A(k)=A\left(\vec{p}(k),k\right)
=\left[\begin{array}{ccc}
\frac{\frac{\partial f_{1}}{\partial p_1}\left(\vec{p}(k),k\right)}{\frac{z(k)}{\sum_{j=1}^n r_j}} & \cdots & \frac{\frac{\partial f_{1}}{\partial p_n}\left(\vec{p}(k),k\right) }{\frac{z(k)}{\sum_{j=1}^n r_j}} \\
\vdots & \ddots & \vdots  \\
\frac{\frac{\partial f_{n}}{\partial p_1}\left(\vec{p}(k),k\right)}{\frac{z(k)}{\sum_{j=1}^n r_j}} & \cdots & \frac{\frac{\partial f_{n}}{\partial p_n}\left(\vec{p}(k),k\right)}{\frac{z(k)}{\sum_{j=1}^n r_j}}
\end{array}\right].\nonumber
\end{equation}
The global equilibrium state is that
$$\vec{d}(k)=\bm{0}_{n} \Longleftrightarrow  f_1(k)=\cdots=f_n(k)=\frac{z(k)}{\sum_{j=1}^n r_j}.$$
However, the global  equilibrium state is hard to reach because we do not know the traffic distribution
next time.
For $k\in\mathbb{Z}^+$ and $i\in\mathcal{N}$, set
\begin{equation}\label{gtilde}
\tilde{g}_k^i(\vec{a}):=\frac{g_k(\vec{a})}{r_i}\Big/\frac{z(k)}{\sum_{j=1}^n r_j}=\frac{\sum_{j=1}^n r_j}{r_i}\cdot\frac{g_k(\vec{a})}{z(k)},
\end{equation}
and for any CPICH transmit power $\vec{p}$, define the \emph{relative busy-degree} of antenna $i$ by
\begin{eqnarray}\label{frel}
f_{i}^{\rm{rel}}\left(\vec{p},k\right)&&:=f_{i}\left(\vec{p},k\right)\Big/\frac{z(k)}{\sum_{j=1}^n r_j}=\frac{\frac{1}{r_i} \int_{\mathbb{R}^n} g_{k}(\vec{a}) \mathbbm{1}_{\left\{p_i- a_i \geq p_j-a_j,\forall j\right\}} \mathrm{d}\vec{a}}{\frac{z(k)}{\sum_{j=1}^n r_j}}\nonumber\\
&&=\int_{\mathbb{R}^n} \tilde{g}_{k}^i(\vec{a}) \mathbbm{1}_{\left\{p_i- a_i \geq p_j-a_j,\forall j\right\}} \mathrm{d}\vec{a}.
\end{eqnarray}
Set
$$\vec{f}^{\rm{rel}}\left(\vec{p},k\right):=\left(f_1^{\rm{rel}}(\vec{p},k),\ldots,f_n^{\rm{rel}}(\vec{p},k)\right)^{\top}.$$
In Lemma \ref{lem_2}, we first give an estimation for the upper bound of $\vec{d}(k)$. The proof is deferred to the Appendix in the supplementary file.

\begin{lemma}\label{lem_2}
Suppose the traffic density function $g_k(\vec{a})$ is continuous on $\mathbb{R}^n, ~k\in \mathbb{Z}^+$,
the graph $\mathcal{G}$ is strongly connected and the elements of its out-degree matrix are positive.
The CPICH transmit power of  antennas are updated by (\ref{solution1}) and (\ref{equ3}).
Then, for any initial state and any constant $\varepsilon >0$, there exists a constant $0<\gamma\leq1$ and
 an integer $k_0\geq 2$ such that for any $k\geq k_0$,
\begin{equation}\label{diff_i}
\|\vec{d}(k)\|_{\infty} \leq \varepsilon+\sum_{t=k-k_0+1}^{k-1} (1-\frac{n^2-n+2}{n^2}\gamma)^{k-t-1}\|\vec{f}^{\rm{rel}}\left(\vec{p}(t+1),t\right)-\vec{f}^{\rm{rel}}\left(\vec{p}(t+1),t+1\right)\|_{\infty}.
\end{equation}
\end{lemma}

From (\ref{diff_i}), we note that the upper bound of $\|\vec{d}(k)\|_{\infty}$  is affected by $\|\vec{f}^{\rm{rel}}\left(\vec{p}(t+1),t\right)-\vec{f}^{\rm{rel}}\left(\vec{p}(t+1),t+1\right)\|_{\infty}$, $t=k-k_0+1,\dots,k-1$. The result of Lemma \ref{lem_2} can be improved if the temporal and spatial variations of network traffic distribution have well properties; therefore, we present the following Theorem \ref{thm_1} and Proposition \ref{corol_1}. The proof of Theorem \ref{thm_1} is involved and is deferred to the Appendix in the supplementary file.
\begin{thm}\label{thm_1}
Assume that the conditions in Lemma \ref{lem_2} still hold.
If there exist a time $k^*\in\mathbb{Z}^+$ such that  the traffic  distribution satisfying
$$g_k(\vec{a})=\beta_k g_{k^*}(\vec{a}),~\forall \vec{a}\in\mathbb{R}^n,~k\geq k^*,$$
where $\{\beta_k\}_{k\geq k^*}$ is a sequence of positive real numbers, and there exists a constant $0<\gamma<1$ such that all antennas' busy-degrees will reach consensus, that is, for any $i\in \mathcal{N}$, $f_i(k)\to \frac{z(k)}{\sum_{j=1}^n r_j}$ as $k\to \infty$.
\end{thm}


Since the assumption that
$g_k(\vec{a})=\beta_k g_{k^*}(\vec{a})$ for any $\vec{a}\in\mathbb{R}^n,$ and $k\geq k^*$, is strong, we propose a weaker assumption.
Set
$$\Omega_i(k):=\left\{\vec{a}: p_i(k)-a_i\geq p_j(k)-a_j\right\}, ~\forall j\in\mathcal{N}$$
to be the coverage  area of antenna $i$ at sampling period $k$.
Then, we can obtain the following Proposition \ref{corol_1} and the Appendix in the supplementary file involves the proof.
\begin{proposition}\label{corol_1}
Assume that the conditions in Lemma \ref{lem_2} still hold.
Suppose that for any $k\in\mathbb{Z}^+$ and $i\in\mathcal{N}$,
\begin{equation}\label{cd}
\left|\int_{\Omega_i(k+1)} \left[\tilde{g}_k^i(\vec{a})-\tilde{g}_{k+1}^i(\vec{a})\right]\mathrm{d}\vec{a}\right|<\delta,
\end{equation}
where $\delta>0$ is a constant. Then, for any initial state and any constant $\varepsilon >0$, and an integer $k_0$, there exists a constant $0<\gamma\leq1$, such that for any $k\geq k_0$, $i \in \mathcal{N}$,
\begin{equation}\label{cor}
|d_i(k)| <\varepsilon +\frac{n^2\delta}{(n^2-n+2)\gamma}.
\end{equation}
\end{proposition}

\begin{remark}
By the definition (\ref{gtilde}), we note that
\begin{equation*}
\int_{\mathbb{R}^n}\left[\tilde{g}_k^i(\vec{a})-\tilde{g}_{k+1}^i(\vec{a})\right]\mathrm{d}\vec{a} 
=\frac{\sum_{j=1}^n r_j}{r_i}\left[\frac{\int g_k(\vec{a})\mathrm{d}\vec{a}}{z(k)}-\frac{\int g_{k+1}(\vec{a})\mathrm{d}\vec{a}}{z(k+1)}\right]=0.
\end{equation*}
In fact,
$$
\tilde{g}_k^i(\vec{a})-\tilde{g}_{k+1}^i(\vec{a})=\frac{\sum_{j=1}^n r_j}{r_i}\left(\frac{g_k(\vec{a})}{z(k)}-\frac{g_{k+1}(\vec{a})}{z(k+1)}\right),
$$
where $g_k(\vec{a})/z(k)$ is the \emph{relative average traffic density function}. When the sampling interval is small enough, the difference between the relative average traffic density function $g_k(\vec{a})/z(k)-g_{k+1}(\vec{a})/z(k+1)$ is also minor.
Therefore, the condition (\ref{cd}) can be satisfied under certain situations.
\end{remark}

As stated in Step 3 of Section \ref{algorithm},
we introduce another algorithm BFDBA that only uses the approximation of $\frac{\partial f_{i}}{\partial p_i}\left(\vec{p}(k),k\right)$. There followed the convergence analysis of BFDBA.

\begin{lemma}\label{lem_3}
Suppose the traffic density function $g_k(\vec{a})$ is continuous on $\mathbb{R}^n, ~k\in \mathbb{Z}^+$.
the graph $\mathcal{G}$ is strongly connected and the elements of its out-degree matrix are positive.
The CPICH transmit power of antennas are updated by (\ref{solu2}) and (\ref{equ3}).
For any initial state and any constant $\varepsilon >0$, there exists an integer $k_0\geq 2$, a constant $\tau>0$ and a constant $0<\gamma\leq1$
such that for any $k\geq k_0$,
\begin{equation} \label{diff2}
\|\vec{d}^{(2)}(k) \|_{1}<\varepsilon+ \sum_{t=k-k_0+1}^{k-1}\nu^{k-t-1}  \|\vec{f}^{\rm{rel}}\left(\vec{p}(t+1),t\right)-\vec{f}^{\rm{rel}}\left(\vec{p}(t+1),t+1\right) \|_{1},
\end{equation}
where $\nu:= \max_{i,k} \left(1- \frac{\gamma \tau z(k)}{\frac{\partial f_{i}}{\partial p_i}\left(\vec{p}(k),k\right)\sum_{j=1}^{n}r_j }\right).$
\end{lemma}
The proof of Lemma \ref{lem_3} is shown in the Appendix in the supplementary file.

Similar to Theorem \ref{thm_1} and Proposition \ref{corol_1}, we can obtain the following Theorem \ref{thm_2} and Proposition \ref{corol_2}. Their proofs are involved in the Appendices in the supplementary file.

\begin{thm}\label{thm_2}
Assume that the conditions in Lemma \ref{lem_3} still hold.
If there exist a time $k^*\in\mathbb{Z}^+$ such that  the traffic  distribution satisfying
$$g_k(\vec{a})=\beta_k g_{k^*}(\vec{a}),~\forall \vec{a}, k\geq k^*,$$
where $\{\beta_k\}_{k\geq k^*}$ is a sequence of positive real numbers,
there exists a constant $0<\gamma\leq1$ and a constant $\tau>0$ such that
for any $i\in \mathcal{N}$, $$f_i(k)\to \left(1-\tau p_i(k)\right)\frac{z(k)}{\sum_{j=1}^n r_j},~k\to \infty.$$
\end{thm}

\begin{remark}
We note that $\tau>0$ should be a small constant and $\tau p_i(k)\frac{z(k)}{\sum_{j=1}^n r_j}$ is minor.
Therefore, all antennas' busy-degrees will reach approximately consensus.
\end{remark}

\begin{proposition}\label{corol_2}
Assume that the conditions in Lemma \ref{lem_3} and the condition (\ref{cd}) still hold.
Then, there exists a constant $0<\gamma\leq1$ and a constant $\tau>0$, for any initial state and any constant $\varepsilon >0$, and an integer $k_0$ such that for any $k\geq k_0$, $i \in \mathcal{N}$,
\begin{equation}\label{c4}
\|\vec{d}^{(2)}(k)\|_{1} <\varepsilon
+\frac{n\delta}{1-\nu}.
\end{equation}
\end{proposition}

\section{Calculation of network coverage based on MR data and MLPs}\label{coverage}
To ensure the coverage of mobile communication network, we need to calculate the minimum CPICH transmit power of each antenna in (\ref{adjust}) for each sampling period.
As stated in Subsection \ref{NCrate}, traditional propagation models suffer from the drawbacks of complex calculation and low accuracy. To achieve dynamic optimization, we need a fast and more accurate method to calculate the real-time network coverage rate.
In this section, we propose a novel calculation method of network coverage rate based on MR data and MLPs. The detailed algorithm is presented as follows.
\subsection{Calculation of network coverage based on MR data}\label{MR}
First, we obtain the data of the CPICH transmit power $p_1,\dots,p_n$, and corresponding MR data at a certain period (such as one day) in advance. Let $\vec{s}^1,\vec{s}^2,\ldots,\vec{s}^K$ denote the collected $K$ pieces of MR data.
For $\ell =1,\ldots,K$, $\vec{s}^\ell= \left( (I_1^{\ell}, s_1^{\ell}),(I_2^{\ell}, s_2^{\ell}),\dots,( I_{m_{\ell}}^{\ell}, s_{m_{\ell}}^{\ell}) \right)$ contains $m_{\ell}$ received CPICH signal strength values and  $I_{m_{\ell}}^{\ell}$ is $s_{m_{\ell}}^{\ell}$'s corresponding antenna ID.
MR data is preprocessed before coverage calculation as follows.\\
\textbf{Step 1 of preprocessing:}
For each received CPICH signal strength value $s_j^{\ell}$ of $\vec{s}^{\ell}$, we calculate the
attenuation value by
\begin{equation}\label{attenua}
a^{\ell}_j:= p_{I_j^{\ell}} - s_j^{\ell},~j=1,\ldots, m_{\ell},
\end{equation}
where $p_{I_j^{\ell}}$ is the CPICH transmit power of antenna $I_j^{\ell}$ \cite{ts1996}.
Then, we can obtain the Table III.
\begin{table}\label{table1}
\centering
\caption{The calculation results of Step 1 of preprocessing.}
\renewcommand\arraystretch{1.5}
\begin{tabular}{m{2.2cm}<{\centering}|m{6cm}<{\centering}}
\hline
IDs of MR data   & Corresponding cells and attenuation values     \\
\hline
$\vec{s}^1$           & $\left( (I_1^1, a_1^1),\ldots,( I_{m_1}^{1}, a_{m_1}^1) \right)$                                  \\
\hline
$\vec{s}^2$           & $\left( (I_1^2, a_1^2),\ldots,( I_{m_1}^{2}, a_{m_1}^2) \right)$                                  \\
\hline
$\cdots$        & $\cdots$                                       \\
\hline
\end{tabular}
\end{table}
\\
\textbf{Step 2 of preprocessing:}
If two rows of MR data $\vec{s}^k$ and $\vec{s}^\ell$ in Table \uppercase\expandafter{\romannumeral3} satisfy:
\begin{itemize}
\item[(1)] $ \left\{ I_1^k,I_2^k,\dots,I_{m_k}^k \right\} \subseteq \left\{ I_1^{\ell},I_2^{\ell},\dots,I_{m_{\ell}}^{\ell} \right\}$,

\item[(2)] if $I_{i}^k = I_j^\ell$, then $a_i^k \geqslant a_j^\ell$,
\end{itemize}
then we regard $\vec{s}^{\ell}$ as a piece of redundant data and we delete the whole row of $\vec{s}^{\ell}$ from Table III. We repeat this process until there is no redundant data.
Let $\vec{s}^1,\vec{s}^2,\ldots,\vec{s}^{K'}$ be the processed MR data and $K'$ is the total.
\\
\textbf{Step 3 of preprocessing:}
If a piece of MR data contains antenna $i$, we say it is antenna $i$'s relevant MR data.
We extract all antennas relevant MR data separately. Let $\left\{ \vec{s}_i^1,\vec{s}_i^2,\ldots,\vec{s}_i^{k_i}\right\} \subseteq  \left\{ \vec{s}^1,\vec{s}^2,\ldots,\vec{s}^{K'} \right\} $ be the set of antenna $i$'s relevant MR data and $k_i$ is its total.
We rank $ \vec{s}_i^1,\vec{s}_i^2,\ldots,\vec{s}_i^{k_i}$ according to their attenuation values with respect to antenna $i$:
$$ a(\vec{s}_i^{1},i) \leqslant  a(\vec{s}_i^2,i)  \leqslant \cdots \leqslant  a(\vec{s}_i^{k_i},i),$$
where $a(\vec{s}_i^{k},i)= p_i -s(\vec{s}_i^{k},i)$, and $s(\vec{s}_i^{k},i)$ is  antenna $i$'s corresponding received CPICH signal strength in $\vec{s}_i^{k},~k=1,\dots,k_i.$
Therefore, we can obtain Table \uppercase\expandafter{\romannumeral4} for antenna $i$.

\begin{table}\label{table2}
\centering
\caption{The ranked  attenuation values of antenna $i$}
\renewcommand\arraystretch{1.5}
\begin{tabular}{m{3.9cm}<{\centering}|m{3.9cm}<{\centering}}
\hline
Attenuation values      & MR data IDs      \\
\hline
$a(\vec{s}_i^{1},i)$      & $\vec{s}_i^{1}$  \\
\hline
$a(\vec{s}_i^2,i) $      & $\vec{s}_i^{2}$  \\
\hline
$\cdots$                & $\cdots$         \\
\hline
\end{tabular}
\end{table}

It is worth noting that the preprocessing of MR data can be done off-line.
With the preprocessed MR data, we introduce the detailed calculation algorithm.\\
\textbf{Step 1:} Assign an initial value ``$0$" to each MR data like Table \uppercase\expandafter{\romannumeral5}.
\begin{table}
\centering
\caption{MR data IDs and their assigned values.}
\renewcommand\arraystretch{1.5}
\begin{tabular}{m{3.9cm}<{\centering}|m{3.9cm}<{\centering}}
\hline
IDs of MR data   &  Assigned Values     \\
\hline
$\vec{s}^1$            &  $0$      \\
\hline
$\vec{s}^2$            &  $0$      \\
\hline
$\cdots$         & $\cdots$    \\
\hline
\end{tabular}
\end{table}
\\
\textbf{Step 2 :} For antenna $i$, calculate its threshold of the attenuation value which satisfies the coverage requirement. Specifically, we denote the threshold by $a_c^i$ and calculate $a_c^i$ by
$$a_c^i := p_i - r_c,$$
where $r_c$ is the threshold of received CPICH signal strength value for users.
For example, the value of $r_c$ can be chosen as $-90$ dbm. If the attenuation value in Table \uppercase\expandafter{\romannumeral5} is less than or equal to the threshold $a_c^i$,
that is, $a(\vec{s}_i^{k},i)\leqslant a_c^i$, we update the assigned value of $\vec{s}_i^{k}$ to ``$1$". Then, we denote the set of MR data whose assigned values are ``$0$" by $\mathcal{S}_f$.
\\
\textbf{Step 3 :} Calculate $F\left(p_1,p_2,\dots,p_n\right)$ by
$$F\left(p_1,p_2,\dots,p_n\right) := 1- \frac{|\mathcal{S}_f|}{K'}.$$

Following these steps, we can calculate the network coverage rate with MR data more accurately.
However, this method may be time-consuming with large volume of MR data.
We analyze the computational complexity of the method.
For Step 1, the assignment is operated over all processed MR data and its complexity is $O(K')$.
Similarly, the complexity for Step 2 is $O(K')$. The computation time of Step 3 can be omitted.
Therefore, the complexity of the total algorithm is $O(K')$,  which could be very demanding for large $K'$, as $K'$ is on the order of $10^8$ in the simulations.
Further, we design a faster calculation method in Subsection \ref{deep}.

\subsection{Fast coverage calculation with MR data and MLPs}\label{deep}
The MLPs (multi-layer perceptrons) are commonly used for function approximation \cite{Pinkus1999,Attali1997}.
Since we can obtain large sets of the CPICH transmit power $p_1,p_2,\dots,p_n$ and calculate the network coverage rate $F\left(p_1,p_2,\dots,p_n\right)$ with the method presented in Subsection \ref{MR},
we can pre-model the complex relationship between input (the CPICH transmit power) and output (the network coverage rate)
with a MLP.
Then, we can calculate the network coverage rate faster using only the CPICH transmit power.

\begin{figure}[htbp]
\centering
\includegraphics[height=2.2in]{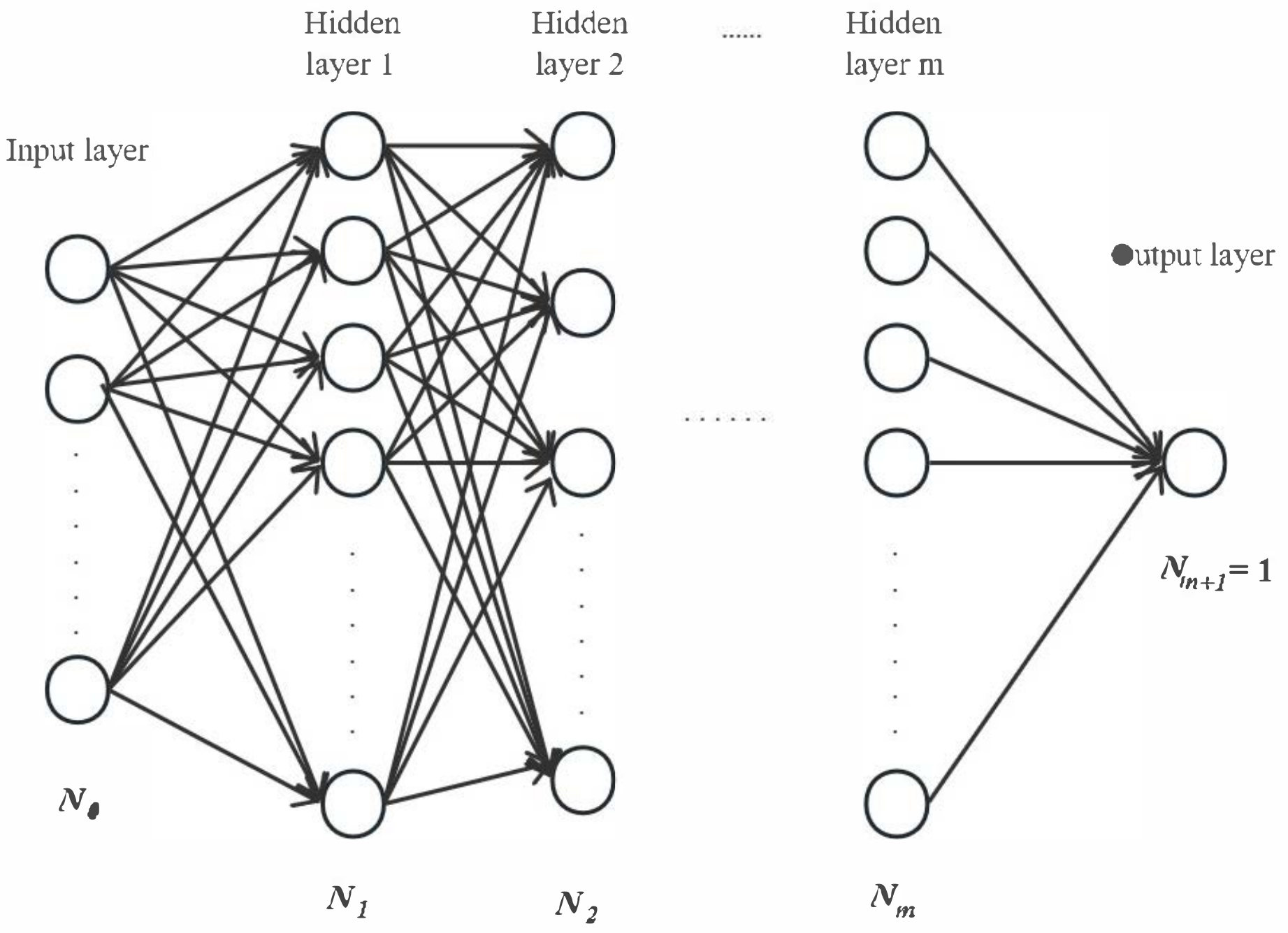}
\caption{The structure of the multi-layer perceptrons.}
\label{Fig:dnn}
\end{figure}

Here, we assume that antenna $i$ and antenna $j$ are neighbours if they appear in the same piece of MR data.
We denote the set of antenna $i$'s neighbours by $\mathcal{N}_i = \{A_1^i,\dots,A_{n_i}^i\}$, where $n_i=|\mathcal{N}_i|$. To simplify the expression, we denote antenna $i$ by $A_0^i$. 
Let $\mathcal{R}_i$ be the cells of antenna $i$ and $\mathcal{N}_i$, and we call $\mathcal{R}_i$  the neighbourhood of antenna $i$.
Besides, we assume that neighbours can communicate directly.
We denote the network coverage rate of $\mathcal{R}_i$ by $F_i(p_{A_0^i},\dots,p_{A_{n_i}^i})$ and we can calculate $F_i$ with the method in Section \ref{MR}. Then, we use $p_{A_j^i},~j=0,\dots,n_i$ as the inputs and $F_i(p_{A_0^i},\dots,p_{A_{n_i}^i})$ as the output to train a MLP.
As we mentioned in Section \ref{description}, when the CPICH transmit power of any antenna is increased, the network coverage rate of its neighbourhood will increase, that is, $F_i$ is a monotone increasing function with respect to $p_{A_0^i},\dots,p_{A_{n_i}^i}$.
Therefore, we make some minor adjustments to the parameters of the classic MLP, transforming it into a monotone increasing model.
The detailed steps for building and training the MLP are presented as follows.\\
\textbf{Step 1:} Normalize the CPICH transmit power $p_{A_j^i},~j=0,\dots,n_i$ by
$$p^*_{A_j^i} = \frac{p_{A_j^i}- p_{\text{min}}(A_j^i)}{p_{\text{max}}(A_j^i)- p_{\text{min}}(A_j^i)},$$
where $p_{\text{max}}(A_j^i)$ and $p_{\text{min}}(A_j^i)$ denote the maximum and minimum CPICH transmit power of $A_j^i$ respectively. Then, $p^*_{A_j^i}$ is the normalized CPICH transmit power of $A_j^i$.\\
\textbf{Step 2:} Establish the MLP and set parameters.
In our application scenario, the number of the input neurons is $n_i +1$ and the number of the output neuron is $1$.
Following the common form \cite{Pinkus1999,Attali1997}, the structure and corresponding parameters of the MLP are shown in Fig. \ref{Fig:dnn}.
To make the MLP be a monotone increasing network, we set the weight between the $p$-th neuron of the $(\ell-1)$-th layer and the $q$-th neuron of the $\ell$-th layer as $(\omega_{qp}^{\ell})^2$, and the bias of the $q$-th neuron of the $\ell$-th layer as $b_q^{\ell}$. We choose the sigmoid function $\sigma(x) = \frac{1}{1+\exp(-x)}$ as the activation function of hidden layers. Then, the monotonicity can be guaranteed. We initialize all parameters to small random values.
\\
\textbf{Step 3:}
We use the mean squared error (MSE) as the cost function to train the parameters with error propagation technique.
Let $W_{(i)}$ and $\vec{b}_{(i)}$ denote the well-trained parameters with respect to antenna $i$.
These parameters can be stored in the memory of antenna $i$.

It is worth noting that the building and training of MLP  can be
done off-line.
With the communication between neighbours, antenna $i$ can obtain the CPICH transmit power $p_{A_1^i},\dots,p_{A_{n_i}^i}$.
With well-trained $W_{(i)}$ and $\vec{b}_{(i)}$ we denote the output by  $F_i^{\star}$.
Therefore, each antenna can calculate the network coverage rate of its neighbourhood fast and accurately in its digital signal processor.

\begin{remark}\label{rem-complex}
It is clear that the computational complexity  of  $F_i^{\star}$  by the MLP is   $O(\sum_{l=0}^{m}N_{\ell}N_{\ell+1})$, where $N_{\ell}$ is the
 number of neurons in hidden layer $\ell$, and $m$ is the number of hidden layers. In our simulations
we choose $m=3$ and the value of $N_{\ell}$ is between $75$ and $3000$, and the computational complexity of $F_i^{\star}$ is lower than
that of  the method presented in Subsection \ref{MR}. Thus, our simulations adopt the MLP method to reduce computation time.
\end{remark}

Based on this fast network coverage rate calculation method, we can further calculate the minimum CPICH transmit power of antenna $i$ in (\ref{adjust}). The calculation process is presented as follows.\\
\textbf{Step 1 :} We say that antenna $i$ fails to satisfy the coverage requirement if $F_i^{\star} < F^{\rm{con}}$. Then, we label such antenna $i$ as $\hat{A}_i$.
Let $\hat{I}$ be the set of labeled antennas' IDs and $\mathcal{V} = \{ \hat{A}_i\}_{i\in \hat{I}}$ be a vertex set. If $\hat{A}_i$ and $\hat{A}_j$ are neighbours, they are regarded as connected and there is an edge between them. Let $\mathcal{E}$ be the edge set, then $\mathcal{G}_f := (\mathcal{V},\mathcal{E}_f)$ is the graph of antennas failing to satisfy the coverage requirement. Let $\{X_h\}_{h=1,\dots,r}$ be the connected components of graph $\mathcal{G}_f$, where $r$ is the  cardinality.
We denote the antenna with the minimum coverage rate in $X_h$ by $A_{X_h}$ and increase its CPICH transmit power by $\Delta p$, where $\Delta p>0$ is a minor adjustable constant. \\
\textbf{Step 2 :} Calculate the new network coverage rate with the well-trained MLP and repeat Step 1 until all antennas satisfy the coverage requirement. At this point, the corresponding CPICH transmit power is the minimum CPICH transmit power in (\ref{adjust}).

With the above steps, all antennas can adjust their CPICH transmit power in the $(k+1)$-th sampling period with (\ref{adjust}).

\section{Simulation Results}\label{simulation}
In this section, we present three sets of simulations based on actual MR and antenna data to verify the effectiveness of our algorithms.
The simulations compare the performance of our algorithms with the current network state, which is the result of the static optimization by base station deployment, and antenna's azimuth and
tilt. The simulation results show that our algorithms perform much better.
In Subsection VII-A, we provide a statement of the datasets. Then, we set up our simulations and present the detailed results and analysis in Subsection VII-B.

\subsection{Dataset Statement}
The datasets used in our simulations are obtained from Beijing Mobile Company. We select three general areas with different antenna densities in Beijing containing both commercial and residential areas, then collect their corresponding one day MR and antenna data, which are referred to as dataset-A, dataset-B, and dataset-C, respectively.

Set the sampling period as one hour and MR data to be classified by hour.
Therefore, the sampling period is $24$, $k = 0,\cdots,23$.
As we stated in Sections \ref{description} and \ref{coverage}, each piece of MR data corresponds to a user in the network and this user accesses the antenna with the highest received strength value.
The dataset-A corresponds to the area shown in Fig. \ref{Fig:map}.
This area includes $1956$ antennas and dataset-A includes the statistics of these antennas' CPICH transmit power and corresponding $112,629,717$ pieces of MR data during the sampling periods.
The dataset-B corresponds to an area with a higher antenna density and includes the statistics of $1658$ antennas' CPICH transmit power and $129,183,183$ pieces of MR data during the sampling periods.
The dataset-C corresponds to a larger area and includes the statistics of $6120$ antennas' CPICH transmit power and $266,485,384$ pieces of MR data during the sampling periods.

\subsection{Simulation setup}

We perform our simulations on a computer platform with Intel i7 CPU (2.9 GHz). The simulation platform is Octave-6.1.0 and we mainly use the svd() built-in function to calculate (\ref{pseudo}).
In the $k-$th sampling period, let $\mathcal{M}(k)$ be the set of MR data of all antennas, and
$\mathcal{M}_i(k)$ be the set of MR data accessing to antenna $i$, $k=0,\cdots,23,~i = 1,\cdots,n$.
Due to the lack of data from the traffic channels, we calculate the busy-degrees based on the number of MR data as a substitute for (\ref{networkload2}).
According to the Law of Large Numbers and the self-similar nature of network traffic \cite{Leland1994}, the average traffic of each antenna is approximately proportional to its accessing user number.
Hence, the average PRB utilization rate of each antenna is approximately proportional to its accessing user number, and our simplification should be rational.
Also, we assume all antennas have the same total of PRBs, i.e., $r_1=r_2=\cdots=r_n$. Therefore, the busy-degree can be calculated by
\begin{equation*}
f_i(k)\approx
\frac{|\mathcal{M}_i(k)|}{M},~k=0,\cdots,23,~i=1,\dots,n,
\end{equation*}
where $M$ is the maximum access user number of antennas.
We say an antenna is \textit{over-busy} if its busy-degree is greater than or equal to $0.7$.
We calculate the standard deviations of all antennas' busy-degrees by $\sqrt{\frac{1}{n}\sum_{i=1}^{n} (f_i(k)-\bar{f}_i(k))^2}$.
Moreover, the volumes of MR data are exceedingly large, making the processing time-consuming.
According to the Large Deviation Principle\cite{touchette2009large},
we perform random sampling on the MR data.
Let $V=30,000,000$ be the sampling scale which can balance computational efficiency and data representativeness.

\begin{figure*}[!t]
\centering
\subfloat[]{\includegraphics[width=2.1in]{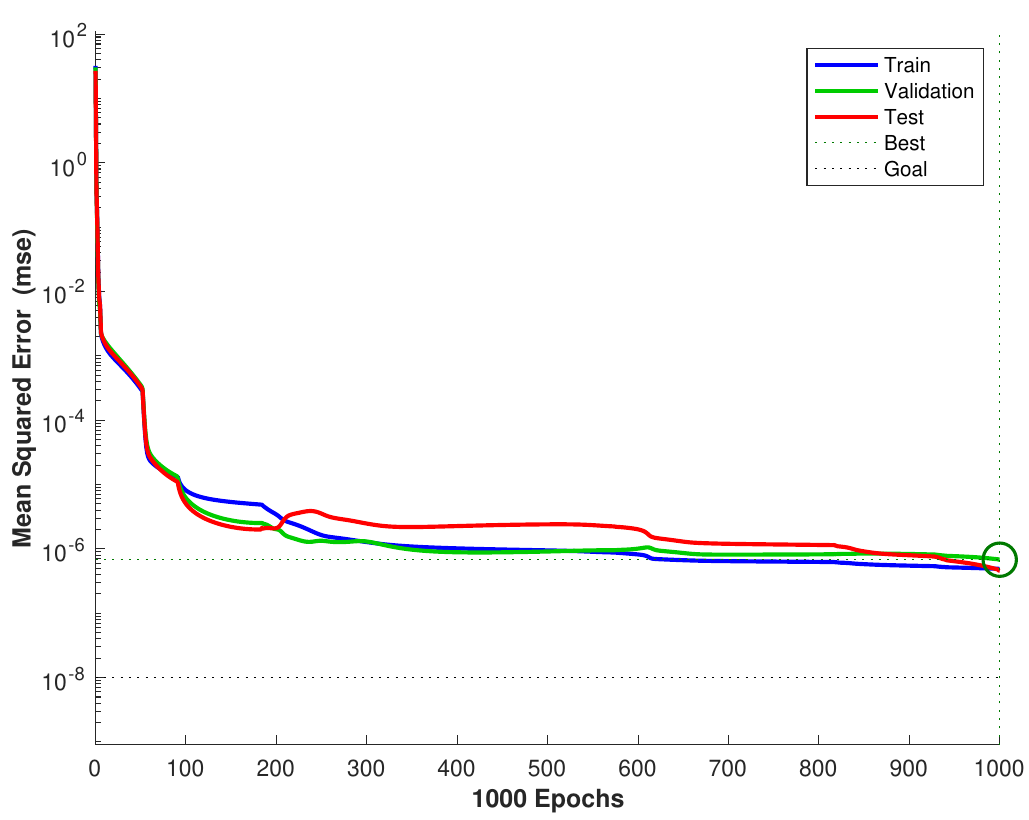}%
\label{performance1}}
\subfloat[]{\includegraphics[width=2.1in]{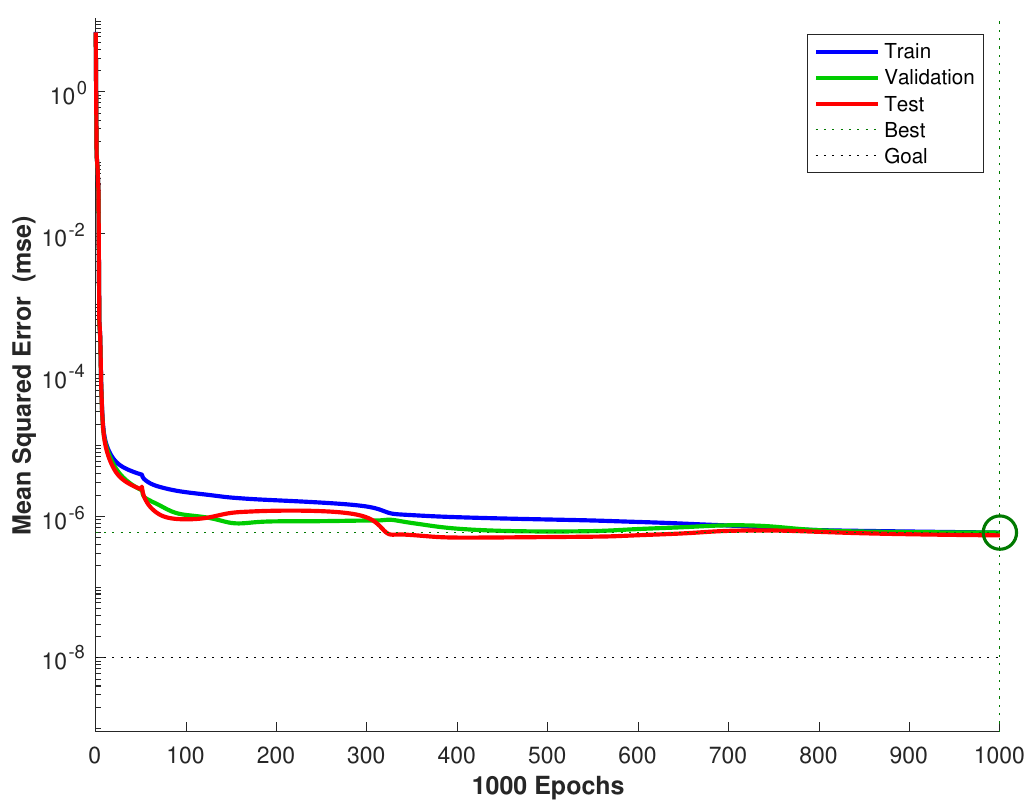}%
\label{performance2}}
\subfloat[]{\includegraphics[width=2.1in]{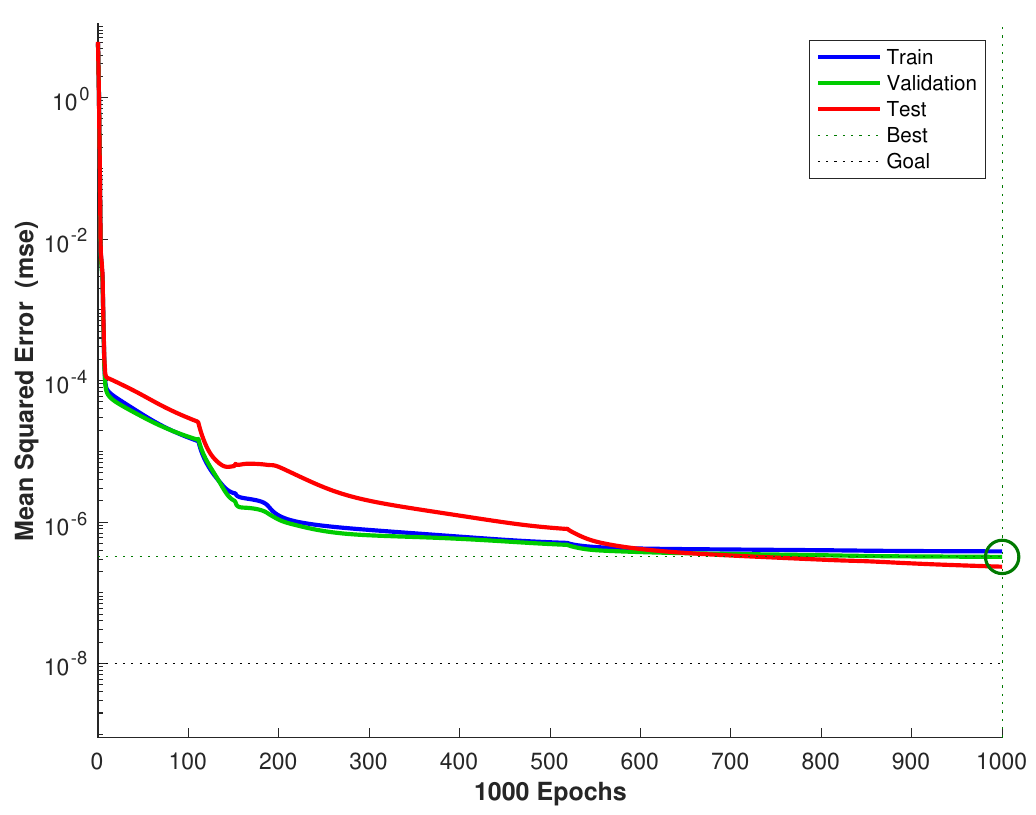}%
\label{performance3}}
\caption{The performances of the MLP corresponding to dataset-A (a); dataset-B (b); and dataset-C (c).}
\label{Fig:mlp}
\end{figure*}

The simulation parameters of our algorithms are summarized as follows.
In simulations of dataset-A,
the adjustable constant used in $A(k)$'s approximation is set to be $\epsilon=0.1$.
For simplification, we do not focus on the tradeoff between network balance and user handover here; this analysis will be addressed in our future research.
Hence, the smoothing factor $\gamma$ is set to be $\gamma = 1$.
The adjustable constant used in $\vec{u}^{(2)}(k)$'s calculation of BFDBA is set to be $\tau = 0.01$.
The minor adjustable constant is set to be $\Delta p =1$.
The maximum access user number of antennas is set to be $M=4000$.
In simulations of dataset-B,
the parameters $\epsilon$, $\gamma$, $\tau$ and $\Delta$ remain unchanged and
the maximum access user number of antennas is set to be $M=5000$.
In simulations of dataset-C,
the parameters $\epsilon$, $\gamma$, $\tau$ and $\Delta$ remain unchanged and
the maximum access user number of antennas is set to be $M=3500$.
From the perspective of global optimization, the target busy-degree of antenna $i$ is calculated by
$$\bar{f_i}(k)=\frac{1}{n}\sum_{j=1}^{n}f_j(k),~i=1,\dots,n.$$
The minimum requirement of network coverage rate $F^{\rm{con}}$ is set as $99.9\%$ and the maximum CPICH transmit power $p_i^{\max}$ is set as $80 ~\mathrm{W}$ for all antennas.
For the setup of the MLPs corresponding to three areas, the number of hidden layers is set to be $3$. For the MLP corresponding to dataset-A, the numbers of neurons in hidden layers are set to be $900$, $460$, and $96$ respectively. For the MLP corresponding to dataset-B, the numbers of neurons in hidden layers are set to be $750$, $290$, and $75$ respectively. For the MLP corresponding to dataset-C, the numbers of neurons in hidden layers are set to be $2700$, $1000$, and $150$ respectively.
The performances of the three MLPs are shown in Fig. \ref{Fig:mlp}.
The training time of each MLP is less than $13$ hours on our computer platform with
Intel i7 CPU.


\subsection{Simulation results and analysis}
\begin{figure*}[htbp]
  \centering
    \subfloat[]{
      \includegraphics[width=3.1in]{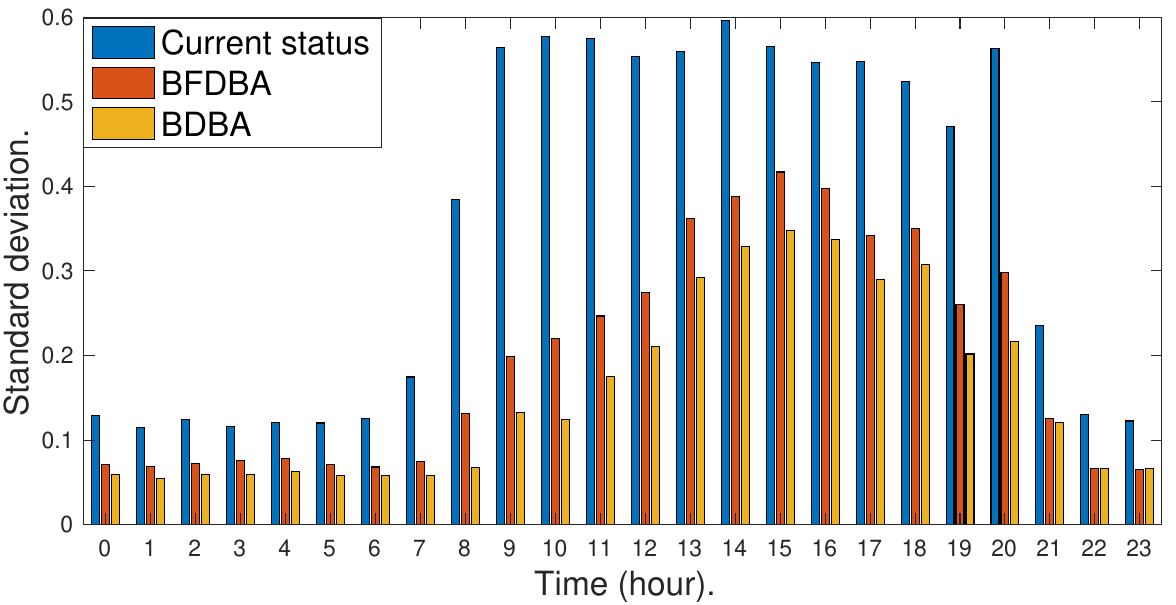}}
      ~~
    \subfloat[]{
      \includegraphics[width=3.1in]{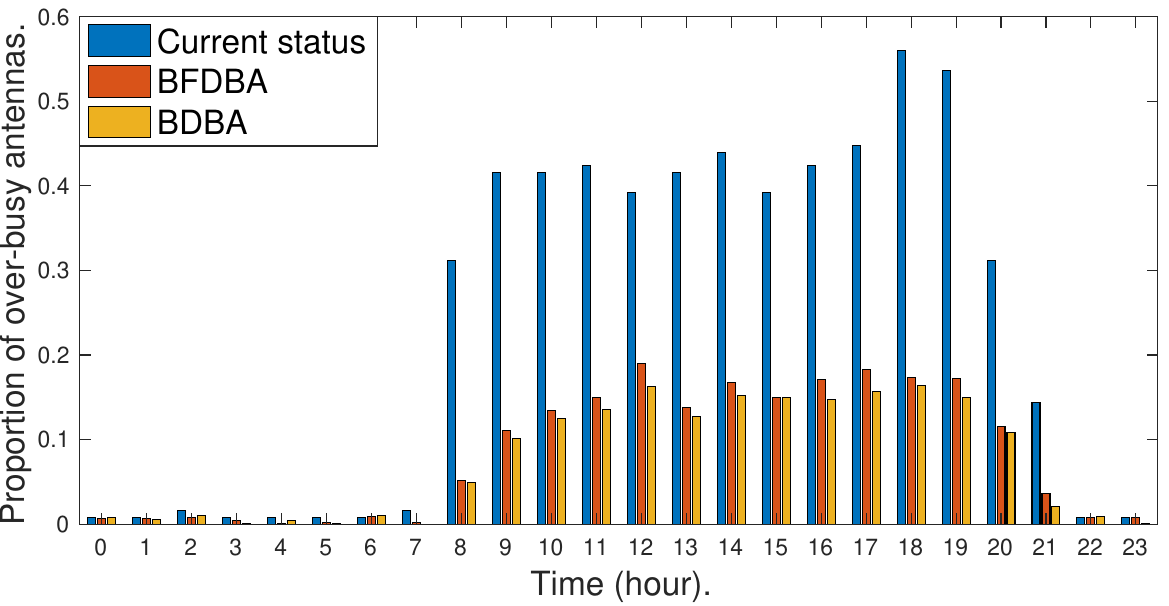}}
\caption{(a) The standard deviations of all antennas' busy-degrees with dataset-A; (b) the proportions of over-busy antennas with dataset-A.}
\label{Fig:control}
\end{figure*}

With dataset-A,
Fig. \ref{Fig:control} shows the standard deviations of all antennas' busy-degrees and the proportions of over-busy antennas. In Fig. \ref{Fig:control}, the blue bars on the left side of the sampling periods correspond to the current network status, the red bars in the middle correspond to the network status with BFDBA, and the yellow bars on the right side correspond to the network status with BDBA.
From Fig. \ref{Fig:control} (a), we can intuitively observe that both BDBA and BFDBA perform well in balancing all antennas' busy-degrees.
To be more specific, compared with the current network status, the mean of the standard deviations of all antennas' busy-degrees is declined by  $55.97\%$ with BDBA; the mean of the standard deviation of all antennas' busy-degrees is declined by $51.02\%$ with BFDBA.
From Fig. \ref{Fig:control} (b), we can obtain that both BDBA and BFDBA effectively reduce the proportions of over-busy antennas.
Compared with the current network status, the mean of the proportions of over-busy antennas is declined by  $68.27\%$ with BDBA; the mean of the proportions of over-busy antennas is declined by  $65.03\%$ with BFDBA.

\begin{figure*}[!t]
\centering
\subfloat[]{\includegraphics[width=2in]{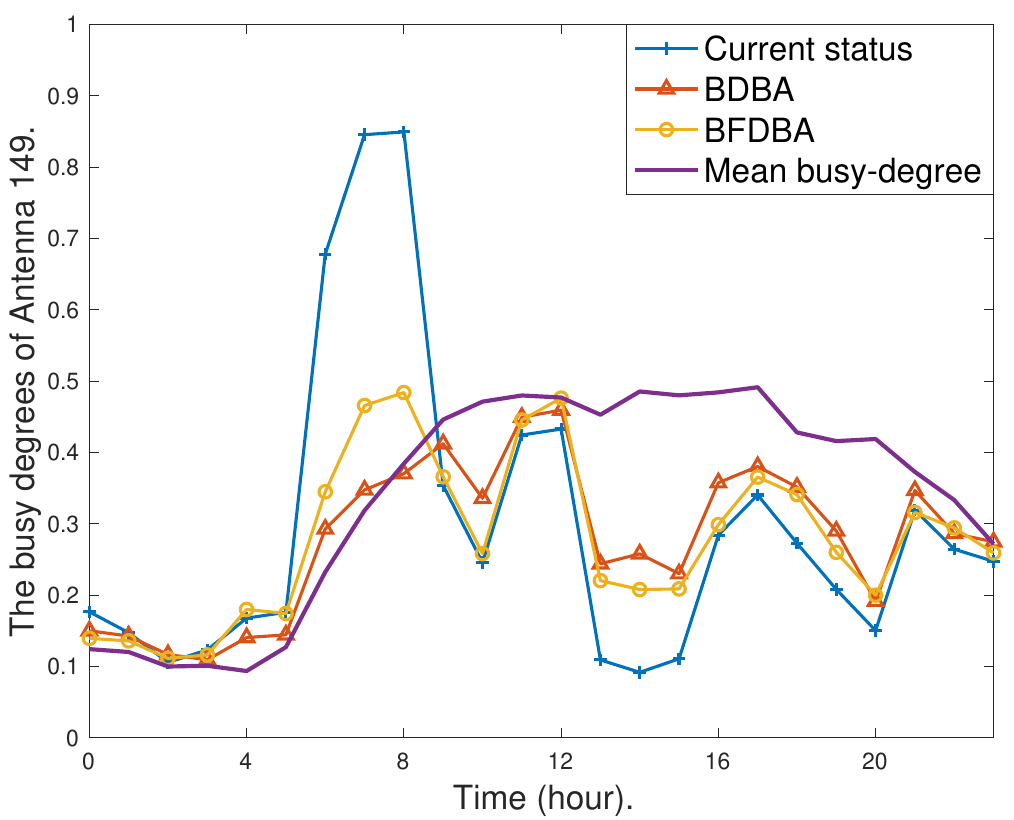}%
\label{select1}}
\subfloat[]{\includegraphics[width=2in]{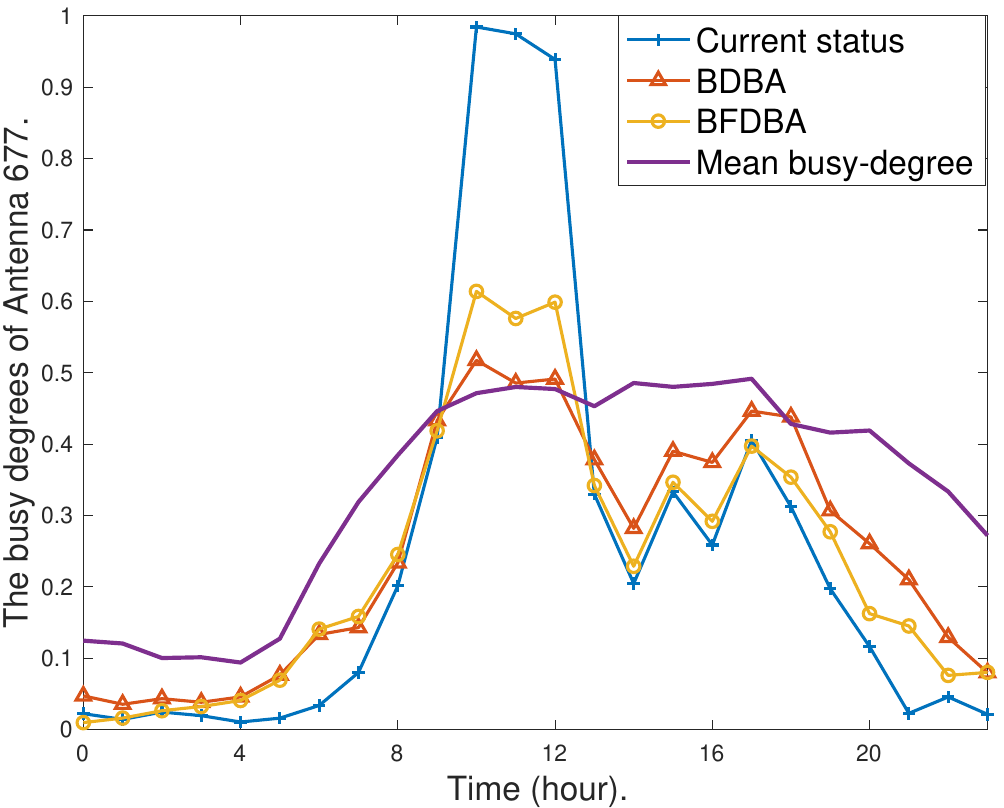}%
\label{select2}}
\\
\subfloat[]{\includegraphics[width=2in]{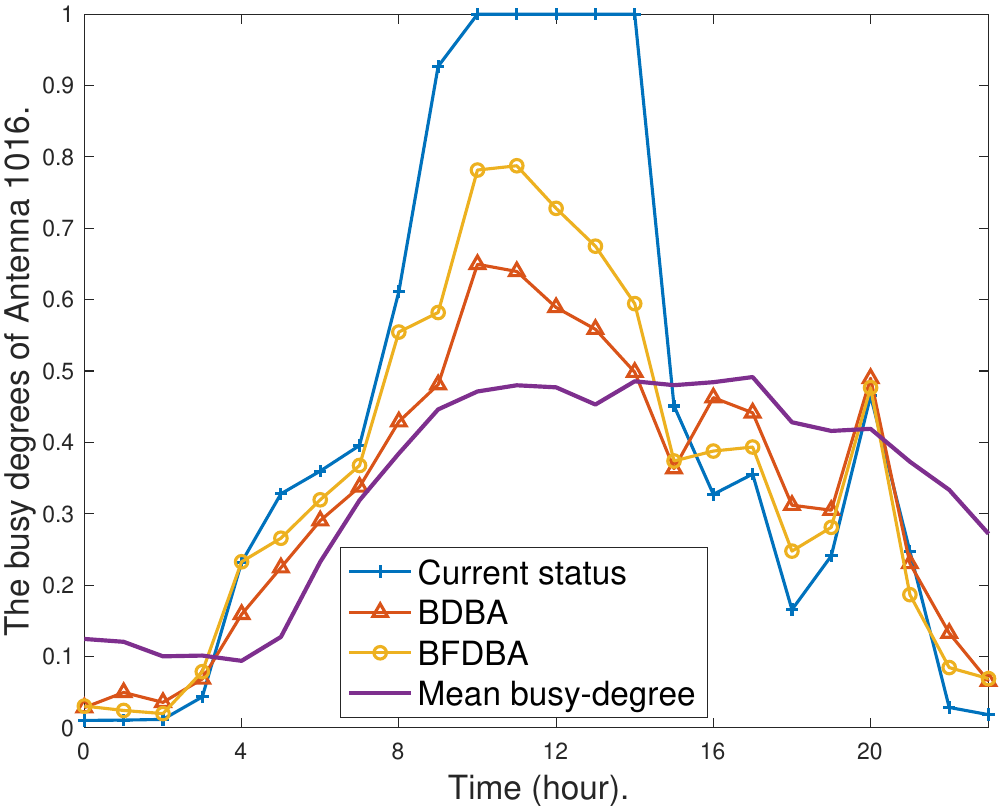}%
\label{select3}}
\subfloat[]{\includegraphics[width=2in]{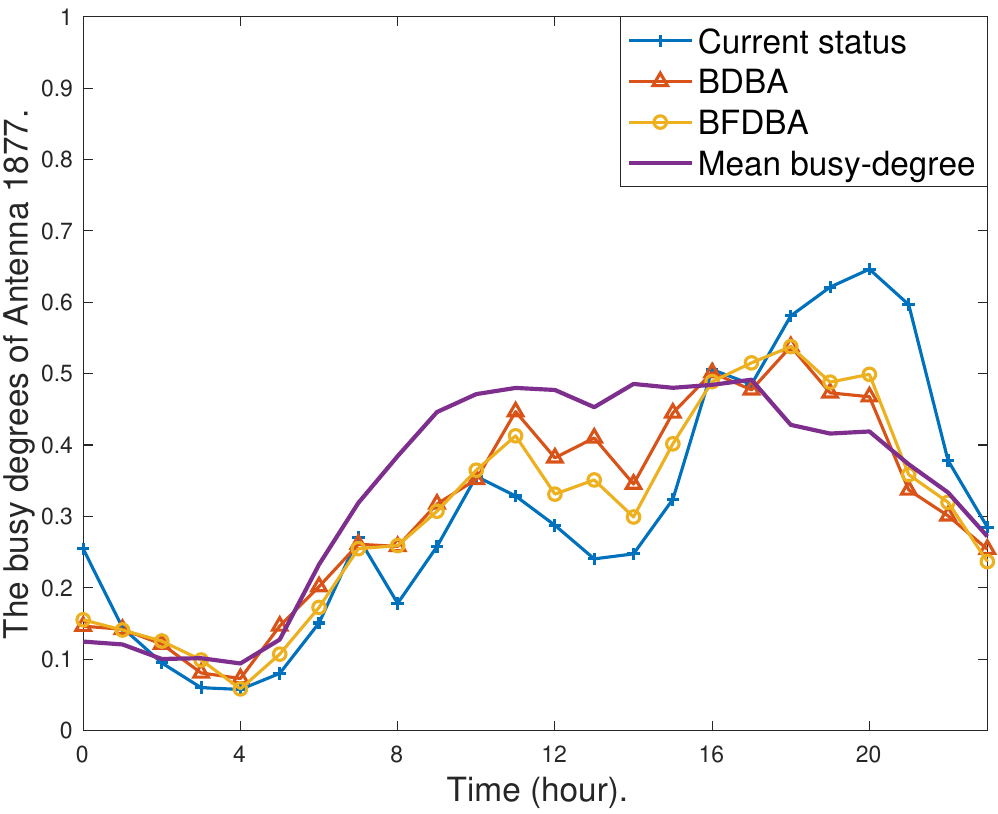}%
\label{select4}}
\caption{The busy-degrees of the randomly selected four antennas under our algorithms. (a) Antenna 149. (b) Antenna 677. (c) Antenna 1016. (d) Antenna 1877.}
\label{Fig:antenna1}
\end{figure*}

Moreover, we specifically compare the busy-degrees of the randomly selected four antennas under our algorithms with the current status.
In the current mobile communication network, the daily changes of selected antennas' busy-degrees are shown in Fig. \ref{Fig:antenna0}.
With BDBA and BFDBA, Fig. \ref{Fig:antenna1} depicts the changes of selected antennas' busy-degrees.
The blue lines with `+' correspond to the current network status; the red lines with `$\triangle$' correspond to  the network status with BDBA; the yellow lines with `o' correspond to  the network status with BFDBA; the smooth solid purple lines correspond to the mean busy-degree of all antennas.
Illustrated by Fig. \ref{Fig:antenna1}, the busy-degrees of the selected antennas get closer to the mean busy-degree of all antennas, indicating that the unbalance of antennas' busy-degrees has been reduced.
Also, the antennas are significantly less busy during their peak periods, since under-utilized antennas can share responsibility for the traffic.
Hence, the number of antennas required for deployment can be reduced, which will help to lower costs of operators.
\begin{figure*}[!t]
  \centering
    \subfloat[]{
      \includegraphics[width=3.1in]{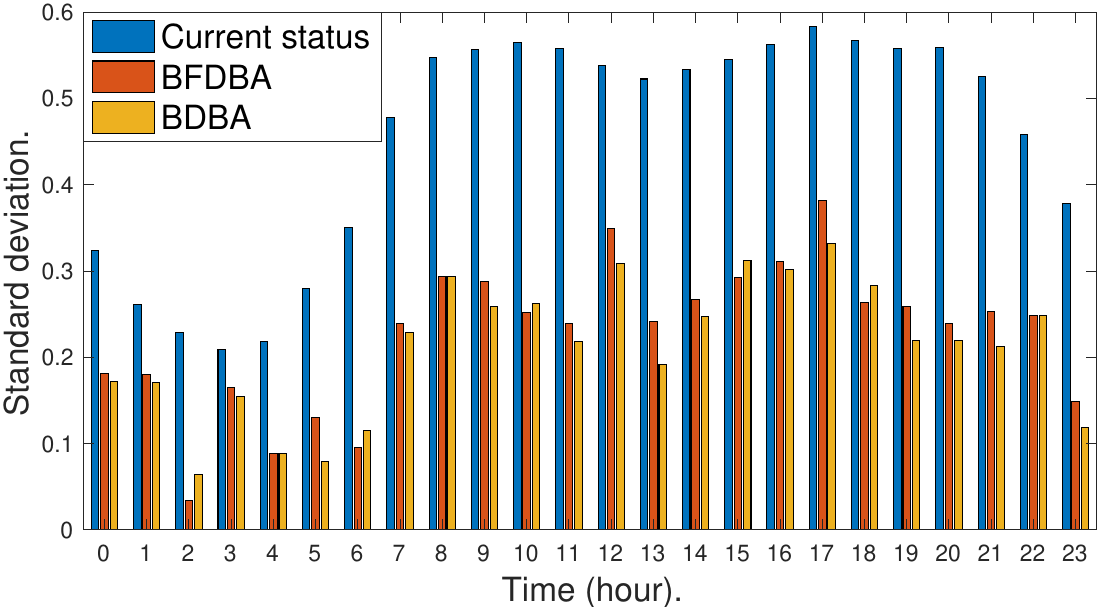}}
      ~~
    \subfloat[]{
      \includegraphics[width=3.1in]{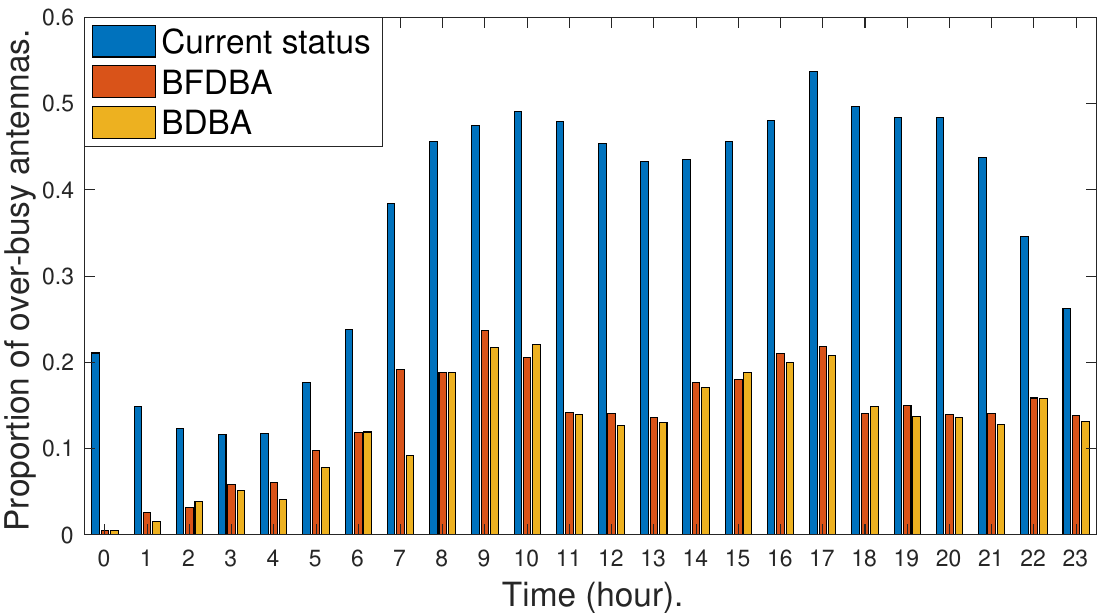}}
\caption{(a) The standard deviations of all antennas' busy-degrees with dataset-B; (b) the proportions of over-busy antennas with dataset-B.}
\label{Fig:control1}
\end{figure*}
\begin{figure*}[!t]
  \centering
    \subfloat[]{
      \includegraphics[width=3.2in]{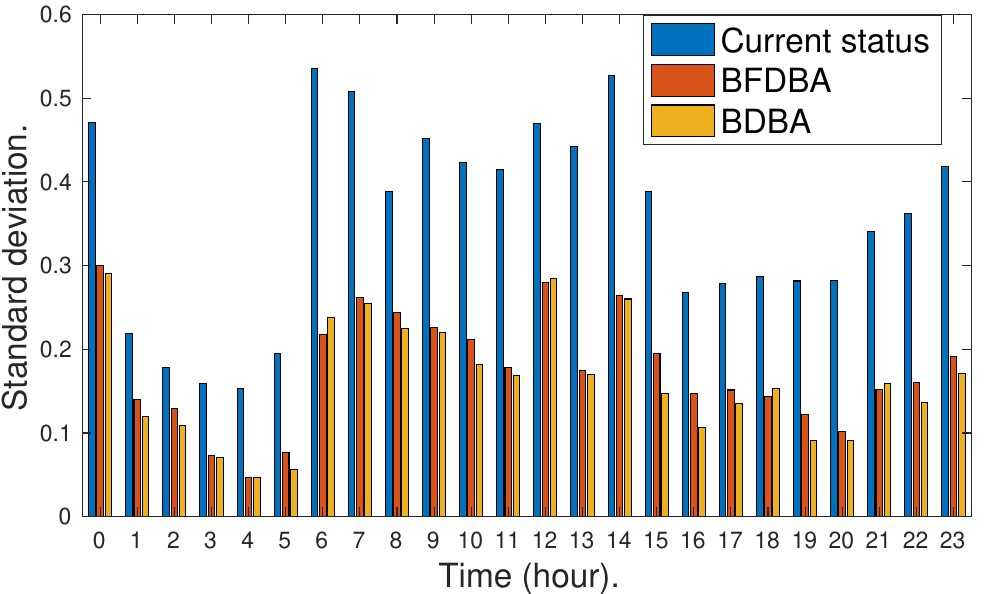}}
      ~~
    \subfloat[]{
      \includegraphics[width=3.2in]{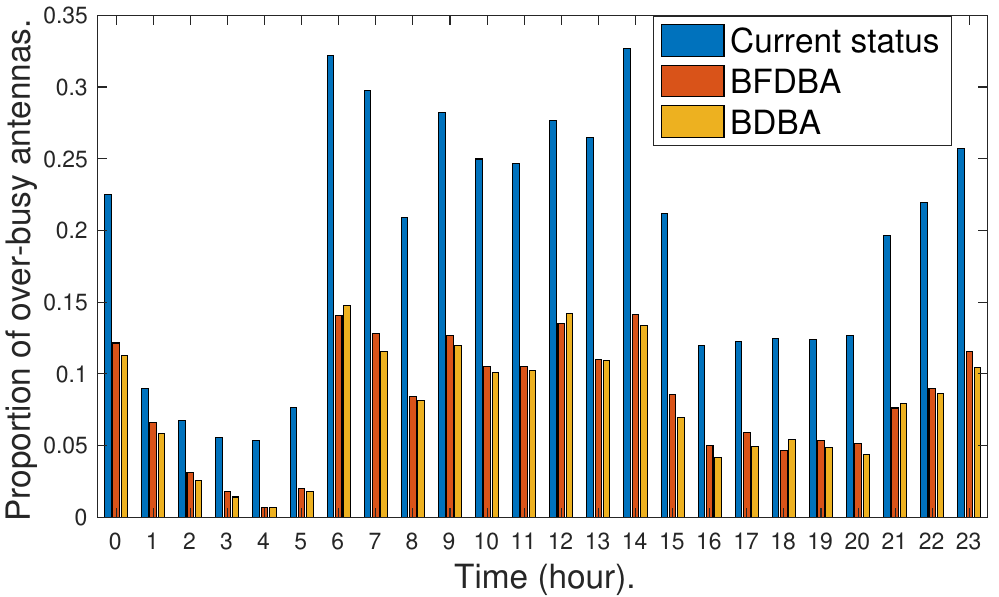}}
\caption{(a) The standard deviations of all antennas' busy-degrees with dataset-C; (b) the proportions of over-busy antennas with dataset-C.}
\label{Fig:control2}
\end{figure*}
Fig. \ref{Fig:control1} and Fig. \ref{Fig:control2} show the simulation results corresponding to the dataset-B and dataset-C, respectively.
Due to the high randomness of user traffic, the performance of our algorithms also exhibit certain fluctuations. Thus, from Fig. \ref{Fig:control}, \ref{Fig:control1} and \ref{Fig:control2}, it can be observed that in some sampling periods BFDBA performs better.
However, the overall performance of BDBA is better. In fact, the mean of the standard deviations of all antennas' busy-degrees is lower under the control of BDBA. To be more specific,
with dataset-B, compared with the current network status, the mean of the standard deviations of all antennas' busy-degrees is declined by $53.78\%$ with BDBA; the mean of the standard deviation of all antennas' busy-degrees is declined by $50.21\%$ with BFDBA; the mean of the proportions of over-busy antennas is declined by $64.74\%$ with BDBA; the mean of the proportions of over-busy antennas is declined by $62.35\%$ with BFDBA.
With dataset-C, compared with the current network status, the mean of the standard deviations of all antennas' busy-degrees is declined by $54.09\%$ with BDBA; the mean of the standard deviation of all antennas' busy-degrees is declined by $50.42\%$ with BFDBA; the mean of the proportions of over-busy antennas is declined by $58.71\%$ with BDBA; the mean of the proportions of over-busy antennas is declined by $56.49\%$ with BFDBA.
The simulations with dataset-B and dataset-C further prove the effectiveness of our algorithms.

\begin{figure*}[!t]
\centering
\subfloat[]{\includegraphics[width=2.1in]{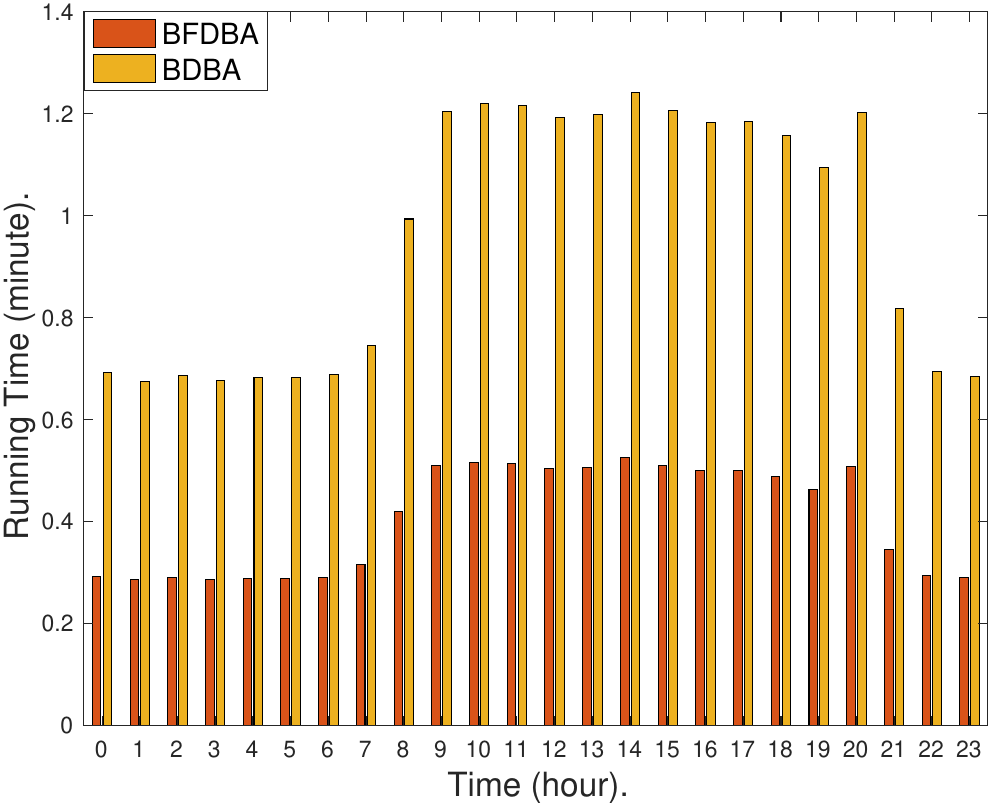}%
\label{time1}}
\subfloat[]{\includegraphics[width=2.1in]{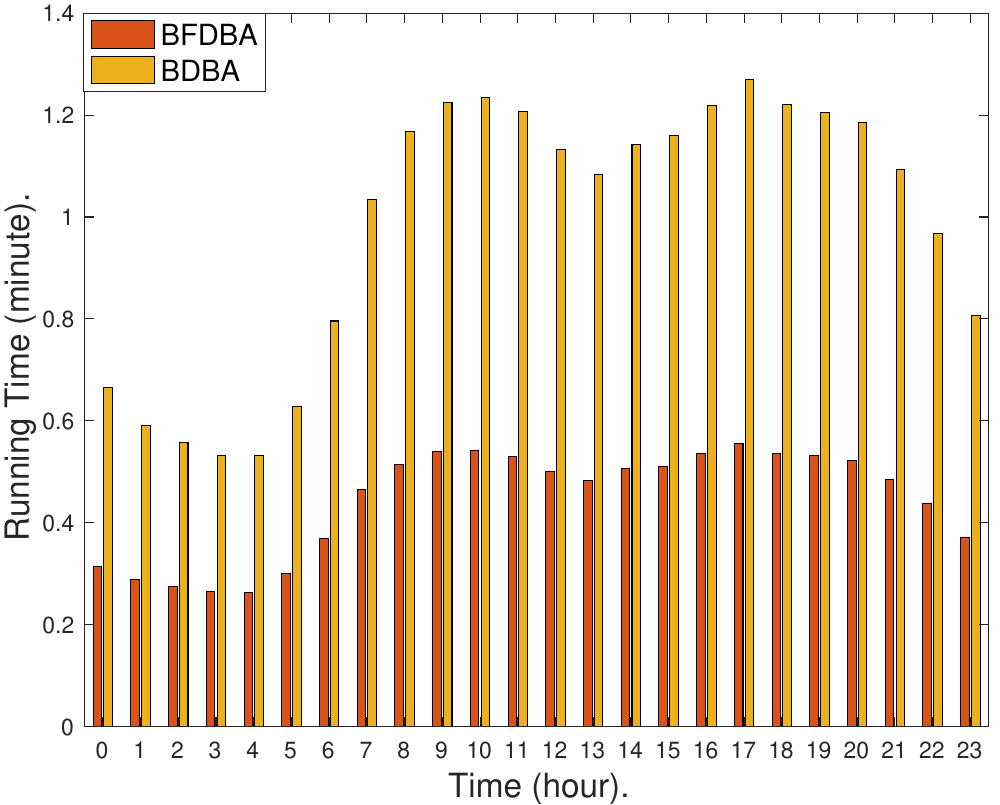}%
\label{time2}}
\subfloat[]{\includegraphics[width=2.1in]{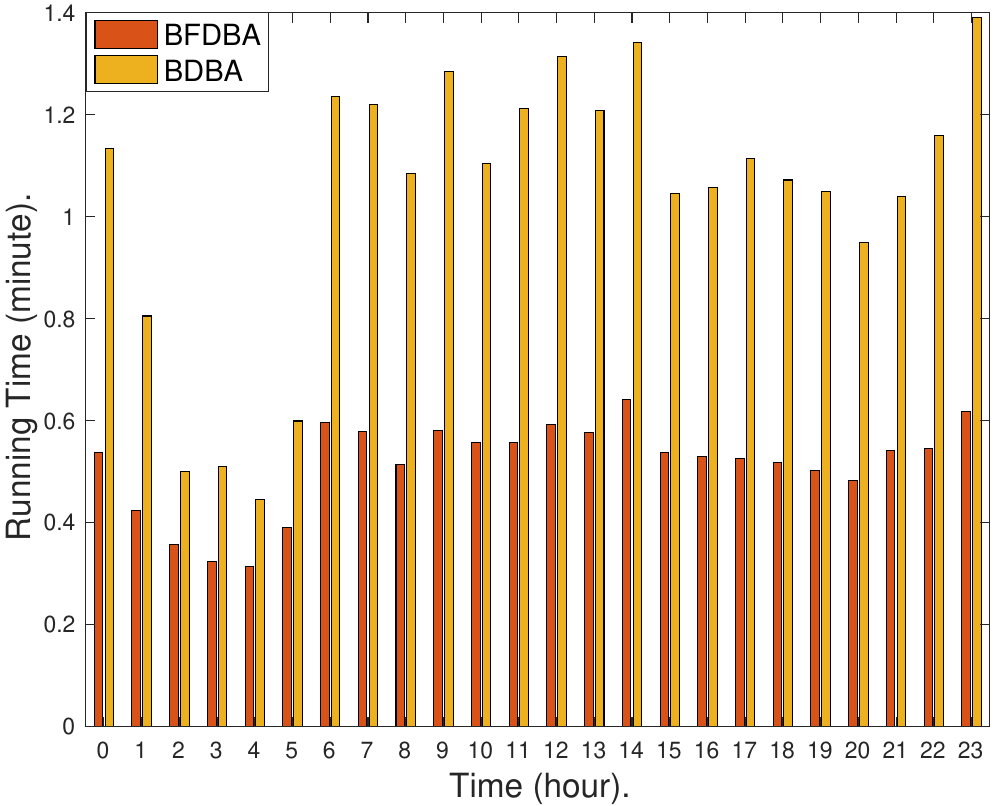}%
\label{time3}}
\caption{The running time comparisons with dataset-A (a);  dataset-B (b); and dataset-C (c).}
\label{Fig:timecomparison}
\end{figure*}

As shown in Fig. \ref{Fig:timecomparison}, we also compare the running time of BDBA and BFDBA. We can observe that BFDBA has an advantage over BDBA as it significantly reduces the running time
in all three sets of simulations. To be more specific, the mean running time of BFDBA is reduced by $52.72\%$ with dataset-A; the mean running time of BFDBA is reduced by $50.59\%$ with dataset-B; the mean running time of BFDBA is reduced by $50.39\%$ with dataset-C.
Therefore, BFDBA can obtain good tradeoff between performance and running time.

\section{Conclusion and Future Work}\label{conclusion}
To optimize the mobile communication network, it is necessary to settle the problem of ``tidal effect''.
However, the dynamic optimization algorithms are still insufficient.
This paper aims to balance the network traffic and proposes two types of dynamic optimization algorithms, BDBA and BFDBA, under the framework of a ``breathing'' mobile communication network.
The busy-degree of each antenna is treated as a multivariate function of all antennas' CPICH transmit power and its Jacobian matrix is approximated with the MR data. Then, the theoretical adjustments of all antennas' CPICH transmit power can be calculated by solving a linear equation. Furthermore, the fast calculation method of network coverage in Section \ref{coverage} can modify the theoretical adjustments to meet the coverage requirement.
Moreover, the theoretical analysis in Section \ref{convergence} proves that the busy-degrees of all antennas can reach the balance (or consensus) under certain assumptions.
As for the simulation results, we observe that the unbalance among antennas is greatly reduced by applying BDBA and BFDBA. The proportions of over-busy antennas are also decreased dramatically.

Since the consensus of all antennas' busy-degrees is affected by traffic distribution, the global equilibrium state is hard to reach.
Therefore, more improvements are needed to enhance the performance of the proposed algorithms.
On one hand, combining the short-term traffic prediction with our algorithms may reduce the impact of variations between sampling periods, which might lead to a more efficient solution.
On the other hand, the proposed algorithms can be combined with the BS sleeping strategies  \cite{liu2015small,Guo2016delay,5992823,5683654} and on-demand coverage strategies \cite{zhao2018deployment}, which can further reduce energy consumption.
In our following work, we will consider this combination and design more refined algorithms.

\appendix
\section{Proof of Lemma 5.1}
We can rewrite (18) as
\begin{equation*}
f_i(k) = \frac{1}{r_i} \int_{-\infty}^{+\infty} \int_{a_i-p_i(k)+p_1(k)}^{+\infty}\cdots \int_{a_i-p_i(k)+p_n(k)}^{+\infty}
g_k(\vec{a})\mathrm{d}a_n \cdots \mathrm{d}a_1  \mathrm{d}a_i,~\forall i \in \mathcal{N}.
\end{equation*}
Since $g_k(\vec{a})$ is continuous on $\mathbb{R}^n, ~k\in \mathbb{Z}^+$, we can exchange the integration order of the repeated integrals.
According to the Leibniz Rule for variable limit integral [36],
for any $j\neq i$, the variable limit integral $\int_{a_i-p_i(k)+p_j(k)}^{+\infty}g_k(\vec{a})\mathrm{d}a_j$ is differentiable with respect to $p_j$.
Hence,
\begin{equation*}
f_i(k) = \frac{1}{r_i} \int_{-\infty}^{+\infty} \int_{a_i-p_i(k)+p_1(k)}^{+\infty}\cdots \int_{a_i-p_i(k)+p_n(k)}^{+\infty}
\left(\int_{a_i-p_i(k)+p_j(k)}^{+\infty}g_k(\vec{a}) \mathrm{d}a_j\right)
 \mathrm{d}a_n \cdots \mathrm{d}a_1  \mathrm{d}a_i
\end{equation*}
is differentiable with respect to $p_j(k)$.
In the same way, for any $j\neq i$, $\int_{a_i-p_i(k)+p_j(k)}^{+\infty}g_k(\vec{a})\mathrm{d}a_j$ is differentiable with respect to $p_i$.
Then, the repeated integral $\int_{a_i-p_i(k)+p_{n-1}(k)}^{+\infty}\int_{a_i-p_i(k)+p_n(k)}^{+\infty}g_k(\vec{a})\mathrm{d}a_n \mathrm{d}a_{n-1}$ is differentiable with respect to $p_i(k)$.
Hence,
\begin{equation*}
f_i(k) = \frac{1}{r_i} \int_{-\infty}^{+\infty} \Bigg(\int_{a_i-p_i(k)+p_1(k)}^{+\infty}\cdots \int_{a_i-p_i(k)+p_n(k)}^{+\infty}
g_k(\vec{a})\mathrm{d}a_n \cdots \mathrm{d}a_1 \Bigg) \mathrm{d}a_i
\end{equation*}
is differentiable with respect to $p_i(k)$.
%
Let $\vec{p}(k):=\left(p_1(k),\ldots,p_n(k)\right).$
By (18), we have
\begin{eqnarray*}
\begin{aligned}
A(k)=A\left(\vec{p}(k),k\right)=
\left[\begin{array}{ccc}
\frac{1}{\bar{f_1}(k)}\frac{\partial f_{1}}{\partial p_1}\left(\vec{p}(k),k\right)  & \cdots & \frac{1}{\bar{f_1}(k)}\frac{\partial f_{1}}{\partial p_n}\left(\vec{p}(k),k\right)\\
\vdots & \ddots & \vdots \\
\frac{1}{\bar{f_n}(k)}\frac{\partial f_{n}}{\partial p_1}\left(\vec{p}(k),k\right)  & \cdots & \frac{1}{\bar{f_n}(k)}\frac{\partial f_{n}}{\partial p_n}\left(\vec{p}(k),k\right)
\end{array}\right].
\end{aligned}
\end{eqnarray*}
By the definition of $g_k(\vec{a})$, we have $g_k(\vec{a}) \geq 0$ for any $\vec{a} \in \mathbb{R}^n$.
On one hand, for all $i \in \mathcal{N}$ and $l\neq i$, we can obtain
\begin{align}
\frac{\partial f_{i}}{\partial p_i}\left(\vec{p}(k),k\right)  
&=\lim_{\Delta p \to 0}\frac{f_{i}\left(\dots,p_i(k)+\Delta p,\dots,k\right)-f_{i}\left(\dots,p_i(k),\dots,k\right)}{\Delta p} \nonumber \\
&=\lim_{\Delta p \to 0}\frac{\frac{1}{r_i}\int_{\mathbb{R}^n} g_k(\vec{a})\cdot \Gamma_{i,i}(k) \mathrm{d}\vec{a}}{\Delta p} \geq 0, \tag{29}
\end{align}
and
\begin{align}
\frac{\partial f_{i}}{\partial p_l}\left(\vec{p}(k),k\right)  
&=\lim_{\Delta p \to 0}\frac{f_{i}\left(\dots,p_l(k)-\Delta p,\dots,k\right)-f_{i}\left(\dots,p_l(k),\dots,k\right)}{-\Delta p} \nonumber \\
&=\lim_{\Delta p \to 0}\frac{\frac{1}{r_i}\int_{\mathbb{R}^n} g_k(\vec{a})\cdot \Gamma_{i,l}(k) \mathrm{d}\vec{a}}{-\Delta p} \leq 0, \tag{30}
\end{align}
where
\begin{align*}
\Gamma_{i,i}(k)&:=
\mathbbm{1}_{\left\{p_i(k)+\Delta p- a_i \geq p_j(k)-a_j,~\forall j\neq i\right\}}
-\mathbbm{1}_{\left\{p_i(k)- a_i \geq p_j(k)-a_j,~\forall j\neq i\right\}},\\
\Gamma_{i,l}(k)&:=\mathbbm{1}_{\left\{p_i(k)- a_i \geq p_j(k)-a_j,\forall j\notin \{i,l\}\right\}} 
\cdot\mathbbm{1}_{\left\{p_i(k)- a_i \geq p_l(k)-\Delta p-a_l\right\}}
-\mathbbm{1}_{\left\{p_i(k)- a_i \geq p_j(k)-a_j,~\forall j\neq i\right\}}.
\end{align*}
On the other hand, we note that $f_i(\vec{p}(k)+\Delta p\bm{1}_n,k)=f_i(\vec{p}(k),k)$ for all $i \in \mathcal{N}$. Hence,
\begin{equation*}
f_i(\vec{p}(k)+\Delta p\bm{1}_n,k)-f_i(\vec{p}(k),k)
=\left(\frac{\partial f_{i}}{\partial p_1}\left(\vec{p}(k),k\right)+\cdots+\frac{\partial f_{i}}{\partial p_n}\left(\vec{p}(k),k\right)\right)\Delta p = 0.
\end{equation*}
Therefore, we have
\begin{equation}
\sum_{j=1}^{n}\frac{\partial f_{i}}{\partial p_j}\left(\vec{p}(k),k\right) = 0,~~i \in \mathcal{N}.\tag{31}
\end{equation}
According to the Definition 6.3 of [32], by (29)-(31), $A(k)$ is a Laplacian matrix.
With the assumption that the elements of $A(k)$'s out-degree matrix are positive, we have
$$\frac{\partial f_{i}}{\partial p_i}\left(\vec{p}(k),k\right)>0.$$
By the Theorem 6.6 of [32], $\mathrm{rank}(A(k))=n-1$ when $\mathcal{G}$ is strongly connected.
$\hfill\blacksquare$

\section{Proof of Lemma 5.2}

First, we calculate the decrease of $d_i(k)$, that is,
\begin{align*}
d_{i}(k+1)-d_{i}(k)&=d_{i}\left(\vec{p}(k+1),k+1\right)-d_{i}\left(\vec{p}(k),k\right)\nonumber\\ \tag{32}
&=d_{i}\left(\vec{p}(k+1),k+1\right)-d_{i}\left(\vec{p}(k+1),k\right)+d_{i}\left(\vec{p}(k+1),k\right)-d_{i}\left(\vec{p}(k),k\right).\nonumber
\end{align*}
By (20), we have
\begin{align*}\label{part2}
d_{i}\left(\vec{p}(k+1),k+1\right)-d_{i}\left(\vec{p}(k+1),k\right)
&= \frac{f_{i}\left(\vec{p}(k+1),k\right)}{\frac{z(k)}{\sum_{j=1}^n r_j}}-\frac{f_{i}\left(\vec{p}(k+1),k+1\right)}{\frac{z(k+1)}{\sum_{j=1}^n r_j}}\nonumber\\
&=f_{i}^{\rm{rel}}\left(\vec{p}(k+1),k\right)-f_{i}^{\rm{rel}}\left(\vec{p}(k+1),k+1\right).\tag{33}
\end{align*}
With the Lagrange mean value theorem [36], (20) and (21), for any $i\in\mathcal{N}$ and $k\in\mathbb{Z}^+$, there exists a constant $\tilde{p}_i = c_i p_i(k)+(1-c_i)p_i(k+1)$ with $c_i\in [0,1],$ such that
\begin{align*}
d_{i}\left(\vec{p}(k+1),k\right)-d_{i}\left(\vec{p}(k),k\right)
=\left(\frac{\partial d_i}{\partial p_1}(\tilde{p},k),\dots,\frac{\partial d_i}{\partial p_n}(\tilde{p},k)\right)\gamma \vec{u}(k)
=-\gamma A_{i,:}(\tilde{p},k)\vec{u}(k),\tag{34}
\end{align*}
where $\tilde{p} = (\tilde{p}_1,\dots,\tilde{p}_n)^\top$, and $A_{i,:}(\tilde{p},k)$ is the $i$-th row of $A(\tilde{p},k)$.
Let $\Sigma(k) = A(\tilde{p},k)-A(k)$.
We note that $\Sigma(k) \to \bm{0}_{n\times n}$ as $\gamma \to 0$, where $\bm{0}_{n\times n}$ is a null matrix.
By (12), we have $\|A^{+}(k)\|_{\infty}\leq \|\Lambda_1^{-1}(k)\|_{\infty} \leq \max_s\frac{1}{\sigma_s(k)}$.
Hence,  by (19), we can obtain
\begin{align*}
-\gamma A_{i,:}(\tilde{p},k)\vec{u}(k) &=-\gamma\left[\left(\Sigma(k)+A(k)\right)A^{+}(k) \right]_{i,:}\vec{d}(k)\nonumber\\
&=-\gamma U_1(k)U_1^{\top}(k)\vec{d}(k)  -\gamma \left[\Sigma(k)A^{+}(k) \right]_{i,:}\vec{d}(k) \nonumber \\
&=-\gamma U_1(k)U_1^{\top}(k)\vec{d}(k)  +o(\gamma), \tag{35}
\end{align*}
where $o(\gamma)$ is said to be infinitesimal compared with $\gamma$.
Together with (11) and (32)-(35), we have
\begin{equation*}
 \vec{d}(k+1)= \left(\bm{I}_n-\gamma U_1(k)U_1^{\top}(k)\right)\vec{d}(k)+o(\gamma)\bm{1}_n
+\vec{f}^{\rm{rel}}\left(\vec{p}(k+1),k\right)-\vec{f}^{\rm{rel}}\left(\vec{p}(k+1),k+1\right),\tag{36}
\end{equation*}
where
\begin{align*}
\bm{I}_n-\gamma U_1(k)U_1^{\top}(k)
=(1-\gamma)\bm{I}_n+\gamma U_2(k)U_2^{\top}(k)=\left[\begin{array}{ccc}
1-\gamma+\gamma\frac{1}{n^2} & \cdots & \gamma\frac{1}{n^2} \\
\vdots & \ddots & \vdots  \\
\gamma\frac{1}{n^2} & \cdots & 1-\gamma+\gamma\frac{1}{n^2}
\end{array}\right],\nonumber
\end{align*}
and $\bm{I}_n$ is the $n$-dimensional identity matrix.
By (36), we can obtain
\begin{equation*}
\|\vec{d}(k+1)\|_{\infty} \leq (1-\frac{n^2-n+2}{n^2}\gamma) \|\vec{d}(k)\|_{\infty}+o(\gamma)
+\|\vec{f}^{\rm{rel}}\left(\vec{p}(k+1),k\right)-\vec{f}^{\rm{rel}}\left(\vec{p}(k+1),k+1\right)\|_{\infty} .
\end{equation*}
Hence,
\begin{align*}
\|\vec{d}(k)\|_{\infty} &\leq (1-\frac{n^2-n+2}{n^2}\gamma)^{k-1}\|\vec{d}(1)\|_{\infty}\\  
&+\frac{n^2}{n^2-n+2}\frac{|o(\gamma)|}{\gamma}+\sum_{t=1}^{k-1} (1-\frac{n^2-n+2}{n^2}\gamma)^{k-t-1} \|\vec{f}^{\rm{rel}}\left(\vec{p}(t+1),t\right)-\vec{f}^{\rm{rel}}\left(\vec{p}(t+1),t+1\right)\|_{\infty}.\nonumber \tag{37}
\end{align*}
Since for any $i\in \mathcal{N},~k \in \mathbb{Z}^+$,
$$\left|f_{i}^{\rm{rel}}\left(\vec{p}(k+1),k\right)-f_{i}^{\rm{rel}}\left(\vec{p}(k+1),k+1\right)\right| < \frac{\sum_{j=1}^nr_j}{r_i},$$
we have that for any  $\varepsilon>0$ and $d_i(1)$, there exists a constant $0<\gamma\leq1$ and
an integer $k_0\geq 2$ such that for any $k> k_0$,
\begin{align*}
&(1-\frac{n^2-n+2}{n^2}\gamma)^{k-1}\|\vec{d}(1)\|_{\infty}+\frac{n^2}{n^2-n+2}\frac{|o(\gamma)|}{\gamma} \nonumber\\
&+\sum_{t=1}^{k-1} (1-\frac{n^2-n+2}{n^2}\gamma)^{k-t-1} \|\vec{f}^{\rm{rel}}\left(\vec{p}(t+1),t\right)-\vec{f}^{\rm{rel}}\left(\vec{p}(t+1),t+1\right)\|_{\infty}<\varepsilon. \tag{38}
\end{align*}
Combining  (37) and (38) yields (23).
$\hfill\blacksquare$

\section{Proof of  Theorem 5.1}
With the condition $g_k(\vec{a})=\beta_k g_{k^*}(\vec{a})$, we can obtain
$$
\tilde{g}_k^i(\vec{a})=\frac{\beta_k g_{k^*}(\vec{a})}{r_i}\Big/\frac{\beta_k z(k^*)}{\sum_{j=1}^n r_j}=\tilde{g}_{k^*}^i(\vec{a}) ,~\forall \vec{a}\in\mathbb{R}^n,~k\geq k^*.
$$
Therefore, by (22), we have
\begin{equation*}
f_{i}^{\rm{rel}}\left(\vec{p}(k+1),k\right)-f_{i}^{\rm{rel}}\left(\vec{p}(k+1),k+1\right) =\int_{\mathbb{R}^n} \left[\tilde{g}_{k}^i(\vec{a})-\tilde{g}_{k+1}^i(\vec{a}) \right]\mathbbm{1}_{\left\{p_i(k+1)- a_i \geq p_j(k+1)-a_j,\forall j\right\}} \mathrm{d}\vec{a}=0. \tag{39}
\end{equation*}
Together with (32)-(35) and (39),
\begin{equation*}
\vec{d}(k+1)-\vec{d}(k) = -\gamma A(\tilde{p},k)\vec{u}(k)=-\gamma\left(\Sigma(k)+A(k)\right)A^{+}(k)\vec{d}(k).\tag{40}
\end{equation*}
By (40), we can obtain
\begin{equation*}
\vec{d}(k+1)=\left[\bm{I}_n-\gamma U_1(k)U_1^{\top}(k)-\gamma\Sigma(k)A^{+}(k)\right]\vec{d}(k), \tag{41}
\end{equation*}
Since $\Sigma(k) \to \bm{0}_{n\times n}$ as $\gamma \to 0$ and $\|A^{+}(k)\|_{\infty}\leq \max_s\frac{1}{\sigma_s(k)}$, there exists $0<\gamma\leq1$ such that
\begin{align*}
\|\bm{I}_n-\gamma U_1(k)U_1^{\top}(k)-\gamma\Sigma(k)A^{+}(k)\|_{\infty} 
&\leq \|\bm{I}_n-\gamma U_1(k)U_1^{\top}(k)\|_{\infty} + \gamma \|\Sigma(k)\|_{\infty}\|A^{+}(k)\|_{\infty}\nonumber\\
&= 1-\frac{n^2-n+2}{n^2}\gamma + o(\gamma)<1. \tag{42}
\end{align*}
Together with (41) and (42), for $k\geq k^*$
\begin{align*}
\|\vec{d}(k+1)\|_{\infty} 
&\leq \|\bm{I}_n-\gamma U_1(k)U_1^{\top}(k)-\gamma\Sigma(k)A^{+}(k)\|_{\infty}  \|\vec{d}(k)\|_{\infty}\\
&\leq \left( 1-\frac{n^2-n+2}{n^2}\gamma + o(\gamma)\right)\|\vec{d}(k)\|_{\infty}.
\end{align*}
Therefore, we have
$$\|\vec{d}(k)\|_{\infty} \leq \left( 1-\frac{n^2-n+2}{n^2}\gamma + o(\gamma)\right)^{k-k^*}\|\vec{d}(k^*)\|_{\infty}.$$
Hence,
$$\lim_{k\to \infty}\|\vec{d}(k)\|_{\infty}= 0,$$
that is, for any $i \in \mathcal{N}$, $f_i(k)\to \frac{z(k)}{\sum_{j=1}^n r_j}$ as $k\to \infty$.
$\hfill\blacksquare$

\section{Proof of Proposition 5.1}
By (22), we have
\begin{equation*}
f_{i}^{\rm{rel}}\left(\vec{p}(k+1),k\right)-f_{i}^{\rm{rel}}\left(\vec{p}(k+1),k+1\right) 
=\int_{\Omega_i(k+1)}\left[\tilde{g}_k^i(\vec{a})-\tilde{g}_{k+1}^i(\vec{a})\right]\mathrm{d}\vec{a}.
\end{equation*}
Together with (24), we can obtain
\begin{equation*}
\left|f_{i}^{\rm{rel}}\left(\vec{p}(k+1),k\right)-f_{i}^{\rm{rel}}\left(\vec{p}(k+1),k+1\right)\right| <\delta. \tag{43}
\end{equation*}
By (43), we have
\begin{align*}
&\sum_{t=k-k_0+1}^{k-1} (1-\frac{n^2-n+2}{n^2}\gamma)^{k-t-1}\|\vec{f}^{\rm{rel}}\left(\vec{p}(t+1),t\right)-\vec{f}^{\rm{rel}}\left(\vec{p}(t+1),t+1\right)\|_{\infty}\nonumber\\
&\leq \sum_{t=1}^{k-1} (1-\frac{n^2-n+2}{n^2}\gamma)^{k-t-1}\delta=\frac{1-(1-\frac{n^2-n+2}{n^2}\gamma)^{k-1}}{\frac{n^2-n+2}{n^2}\gamma}\delta \leq \frac{n^2}{n^2-n+2}\frac{\delta}{ \gamma }. \tag{44}
\end{align*}
Together with (37) and (44), we have
\begin{equation*}
\|\vec{d}(k)\|_{\infty} \leq (1-\frac{n^2-n+2}{n^2}\gamma)^{k-1}\|\vec{d}(1)\|_{\infty}  +\frac{n^2}{n^2-n+2}\frac{|o(\gamma)|}{\gamma} + \frac{n^2}{n^2-n+2}\frac{\delta}{ \gamma }. \tag{45}
\end{equation*}
Then, for any initial state and any constant $\varepsilon >0$, there exists an integer $k_0$ such that for any $k\geq k_0$, $i \in \mathcal{N}$,
\begin{equation*}
 (1-\frac{n^2-n+2}{n^2}\gamma)^{k-1}\|\vec{d}(1)\|_{\infty} \\ +\frac{n^2}{n^2-n+2}\frac{|o(\gamma)|}{\gamma} <\varepsilon. \tag{46}
\end{equation*}
Combining  (45) and (46) yields (25).
$\hfill\blacksquare$

\section{Proof of Lemma 5.3}
Same as the proof of Lemma 5.2, we have
\begin{equation*}
\vec{d}^{(2)}(k+1)-\vec{d}^{(2)}(k) = \vec{d}^{(2)}\left(\vec{p}(k+1),k+1\right)- \vec{d}^{(2)}\left(\vec{p}(k+1),k\right)+\vec{d}^{(2)}\left(\vec{p}(k+1),k\right)-\vec{d}^{(2)}\left(\vec{p}(k),k\right), \tag{47}
\end{equation*}
where
\begin{equation*}
\vec{d}^{(2)}\left(\vec{p}(k+1),k+1\right)- \vec{d}^{(2)}\left(\vec{p}(k+1),k\right) 
=\vec{f}^{\rm{rel}}\left(\vec{p}(k+1),k\right)-\vec{f}^{\rm{rel}}\left(\vec{p}(k+1),k+1\right).
\end{equation*}
By (34) and (17),
\begin{align*}
\vec{d}^{(2)}\left(\vec{p}(k+1),k\right)-\vec{d}^{(2)}\left(\vec{p}(k),k\right)&=-\gamma \left(A(\tilde{p},k)+\tau \bm{I}_n\right)\vec{u}^{(2)}(k)\\
&=-\gamma \left(A(\tilde{p},k)+\tau \bm{I}_n\right)D^{-1}(k)\vec{d}^{(2)}(k),
\end{align*}
Together with (47), we have
\begin{equation*}
\vec{d}^{(2)}(k+1)= \left(\bm{I}_n-\gamma \left(A(\tilde{p},k)+\tau \bm{I}_n\right)D^{-1}(k)\right) \vec{d}^{(2)}(k)
+\vec{f}^{\rm{rel}}\left(\vec{p}(k+1),k\right)-\vec{f}^{\rm{rel}}\left(\vec{p}(k+1),k+1\right), \tag{48}
\end{equation*}
where
\begin{equation*}
\bm{I}_n-\gamma \left(A(\tilde{p},k)+\tau \bm{I}_n\right)D^{-1}(k)= 
\left[\begin{array}{ccc}
1-\gamma \frac{\frac{\partial f_{1}}{\partial p_1}(\tilde{p},k)+\tau \frac{z(k)}{\sum_{j=1}^{n}r_j}}{ \frac{\partial f_1}{\partial p_1}\left(\vec{p}(k),k\right)}  & \cdots & -\gamma \frac{\frac{\partial f_1}{\partial p_n}(\tilde{p},k)}{\frac{\partial f_n}{\partial p_n}\left(\vec{p}(k),k\right)} \\
\vdots & \ddots & \vdots \\
-\gamma\frac{\frac{\partial f_n}{\partial p_1}(\tilde{p},k)}{\frac{\partial f_1}{\partial p_1}\left(\vec{p}(k),k\right)} & \cdots & 1-\gamma \frac{\frac{\partial f_n}{\partial p_n}(\tilde{p},k)+\tau \frac{z(k)}{\sum_{j=1}^{n}r_j}}{ \frac{\partial f_n}{\partial p_n}\left(\vec{p}(k),k\right)}
\end{array}\right].
\end{equation*}
By (6), we have for any $i\in \mathcal{N}$,
$$\sum_{j=1}^n \frac{\partial f_{j}}{\partial p_i}(\tilde{p},k) =\frac{\partial \sum_{j=1}^n f_{j}}{\partial p_i}(\tilde{p},k)=\frac{\partial\frac{nz(k)}{\sum_{j=1}^n r_j}}{\partial p_i}(\tilde{p},k)=0.$$
Hence, there exists $0<\gamma\leq1$ and $\tau>0$ such that
\begin{equation*}
\|\bm{I}_n-\gamma \left(A(\tilde{p},k)+\tau \bm{I}_n\right)D^{-1}(k)\|_{1}  
= \max_{i}\left(1-\gamma \tau \frac{\frac{z(k)}{\sum_{j=1}^{n}r_j}}{\frac{\partial f_{i}}{\partial p_i}\left(\vec{p}(k),k\right)} \right)<1. \tag{49}
\end{equation*}
By (48), we can obtain that
\begin{align*}
\|\vec{d}^{(2)}(k+1) \|_{1} 
&\leq\|\bm{I}_n-\gamma \left(A(\tilde{p},k)+\tau \bm{I}_n\right)D^{-1}(k)\|_{1}\cdot \|\vec{d}^{(2)}(k) \|_{1} +\|\vec{f}^{\rm{rel}}\left(\vec{p}(k+1),k\right)-\vec{f}^{\rm{rel}}\left(\vec{p}(k+1),k+1\right) \|_{1}\\
&\leq \nu \|\vec{d}^{(2)}(k) \|_{1}+\|\vec{f}^{\rm{rel}}\left(\vec{p}(k+1),k\right)-\vec{f}^{\rm{rel}}\left(\vec{p}(k+1),k+1\right) \|_{1}.
\end{align*}
Hence, we have
\begin{equation*}
\|\vec{d}^{(2)}(k) \|_{1}  \leq \nu^{k-1} \|\vec{d}^{(2)}(1) \|_{1}
+\sum_{t=1}^{k-1}\nu^{k-t-1}
\times\|\vec{f}^{\rm{rel}}\left(\vec{p}(t+1),t\right)-\vec{f}^{\rm{rel}}\left(\vec{p}(t+1),t+1\right) \|_{1}.\tag{50}
\end{equation*}
Similar to the proof of Theorem 5.1, by (49), for any $\varepsilon$, there exists an integer $k_0\geq 2$, for any $k> k_0$, such that
\begin{equation*}
\nu^{k-1} \|\vec{d}^{(2)}(1) \|_{1} +\sum_{t=1}^{k-k_0} \nu^{k-t-1}\|\vec{f}^{\rm{rel}}\left(\vec{p}(t+1),t\right)-\vec{f}^{\rm{rel}}\left(\vec{p}(t+1),t+1\right) \|_{1}<\varepsilon.
\end{equation*}
Therefore, we can obtain (26).
$\hfill\blacksquare$

\section{Proof of Theorem 5.2}
Same as the proof of Theorem 5.1, (39) still holds.
Combining (47) and (48), we have
$$
\vec{d}^{(2)}(k+1)= \left(\bm{I}_n-\gamma \left(A(\tilde{p},k)+\tau \bm{I}_n\right)D^{-1}(k)\right)  \vec{d}^{(2)}(k).
$$
Hence,
$$
\|\vec{d}^{(2)}(k+1)\|_{1} \leq \nu\|\vec{d}^{(2)}(k) \|_{1}.
$$
Therefore, we have
$$\|\vec{d}^{(2)}(k)\|_{1} < \nu^{k-k^*}\|\vec{d}^{(2)}(k^*)\|_{1}.$$
Similar to Theorem 5.1, we have
$$\lim_{k\to \infty}\|\vec{d}^{(2)}(k)\|_{1}= 0,$$
that is,
$$\lim_{k\to \infty}\|\vec{d}(k)-\tau\vec{p}(k)\|_{1}= 0.$$
Hence, $f_i(k)\to\left(1-\tau p_i(k)\right)\frac{z(k)}{\sum_{j=1}^n r_j}$ as $k\to \infty$.
$\hfill\blacksquare$

\section{Proof of Proposition 5.2}
Similar to (44) of Proposition 5.1, we have
\begin{align*}
&~~~~\sum_{t=1}^{k-1} \nu^{k-t-1}\|\vec{f}^{\rm{rel}}\left(\vec{p}(t+1),t\right)-\vec{f}^{\rm{rel}}\left(\vec{p}(t+1),t+1\right) \|_{1}\nonumber\\
&=  \sum_{t=1}^{k-1} \nu^{k-t-1}\sum_{i=1}^n|f_{i}^{\rm{rel}}\left(\vec{p}(t+1),t\right)-f_{i}^{\rm{rel}}\left(\vec{p}(t+1),t+1\right)|\nonumber\\
&\leq \sum_{t=1}^{k-1} \nu^{k-t-1} n\int_{\mathbb{R}^n}\delta \mathbbm{1}_{\left\{p_i(k+1)- a_i \geq p_j(k+1)-a_j,\forall j\right\} \mathrm{d}\vec{a}} \nonumber\\
&\leq n\delta \sum_{t=1}^{k-1} \nu^{k-t-1} = \frac{n\delta(1-\nu^{k-1})}{1-\nu}<\frac{n\delta}{1-\nu}. \tag{51}
\end{align*}
By (50) and (51), we have
\begin{equation*}
\|\vec{d}^{(2)}(k) \|_{1}  \leq \nu^{k-1} \|\vec{d}^{(2)}(1) \|_{1}
+\frac{n\delta}{1-\nu}. \tag{52}
\end{equation*}
Then, for any initial state and any constant $\varepsilon >0$, there exists a constant $\gamma$ and an integer $k_0$ such that for any $k\geq k_0$, $i \in \mathcal{N}$,
\begin{equation*}
\nu^{k-1} \|\vec{d}^{(2)}(1) \|_{1} <\varepsilon.\tag{53}
\end{equation*}
Combining  (52) and (53) yields (27). 
$\hfill\blacksquare$



\begin{thebibliography}{1}
\bibliographystyle{IEEEtran}

\bibitem{pomp2016elastic}
D. Pompili, A. Hajisami, and T. X. Tran, ``Elastic resource utilization framework for high capacity and energy efficiency in cloud RAN,'' \emph{IEEE Communications Magazine},
  vol.~54, no.~1, pp. 26--32, 2016.

\bibitem{khan2015reducing}
M. Khan, R. S. Alhumaima, and H. S. Al-Raweshidy,  ``Reducing energy consumption by dynamic resource allocation in C-RAN,'' in \emph{European Conf. on Networks and Communications (EuCNC).}, Paris, France,  2015, pp. 169--174. DOI: 10.1109/EuCNC.2015.7194062.

\bibitem{niu2011tango}
Z. Niu,  ``TANGO: traffic-aware network planning and green operation,'' \emph{IEEE Wireless Communications},
  vol.~18, no.~5, pp. 25--29, 2011.


\bibitem{Xu2019}
Y. Xu, F. Yin, W. Xu, \emph{et~al.}, ``Wireless traffic prediction with scalable Gaussian process: Framework, algorithms, and verification'', \emph{IEEE Journal on Selected Areas in Communications}, vol.~37, no.~6, pp. 1291--1306, 2019.

\bibitem{Nagib2021}
A. M. Nagib, H. Abou-Zeid, H. S. Hassanein, A. Bin Sediq, and G. Boudreau, ``Deep Learning-Based Forecasting of Cellular Network Utilization at Millisecond Resolutions,'' in \emph{IEEE International Conference on Communications}, Montreal, QC, Canada, 2021, pp. 1--6. DOI: 10.1109/ICC42927.2021.9500858.

\bibitem{Murudkar2019}
C. V. Murudkar and R. D. Gitlin, ``Optimal-Capacity, Shortest Path Routing in Self-Organizing 5G Networks using Machine Learning,'' in \emph{IEEE 20th Wireless and Microwave Technology Conference},  Cocoa Beach, FL, USA, 2019, pp. 1--5. DOI: 10.1109/WAMICON.2019.8765434.


\bibitem{Mishra2014}
S. Mishra, R. Thakur, and C. S. R. Murthy, ``An Efficient Physical Resource Block Assignment for Dense Femtocell Networks,'' in \emph{IEEE 79th Vehicular Technology Conference},  Seoul, Korea (South), 2014, pp. 1--5. DOI: 10.1109/VTCSpring.2014.7022819.



\bibitem{huang2019geometric}
X. Huang, S. Tang, D. Zhang, \emph{et~al.}, ``Geometric approach based resource allocation in heterogeneous cellular networks'', \emph{IEEE Transactions on Vehicular Technology},
  vol.~68, no.~12, pp. 11902--11914, 2019.



\bibitem{auer2011much}
G.~Auer, V.~Giannini, C.~Desset, I.~Godor, \emph{et~al.}, ``How much energy
  is needed to run a wireless network?'' \emph{IEEE Wireless Communications},
  vol.~18, no.~5, pp. 40--49, 2011.

\bibitem{cho2013energy}
S. Cho and W. Choi, ``Energy-efficient repulsive cell activation for heterogeneous cellular networks,'' \emph{IEEE Journal on Selected Areas in Communications},
  vol.~31, no.~5, pp. 870--882, 2013.

\bibitem{it2020toward}
I. Tomkos, D. Klonidis, E. Pikasis, and S. Theodoridis, ``Toward the 6G Network Era: Opportunities and Challenges,'' \emph{IT Professional},
  vol.~22, no.~1, pp. 34--38, 2020.





%
%
%


\bibitem{5992823}
K.~Son, H.~Kim, Y.~Yi, and B.~Krishnamachari, ``Base station operation and user
  association mechanisms for energy-delay tradeoffs in green cellular
  networks,'' \emph{IEEE Journal on Selected Areas in Communications}, vol.~29,
  no.~8, pp. 1525--1536, 2011.

\bibitem{5683654}
E.~Oh and B.~Krishnamachari, ``Energy savings through dynamic base station
  switching in cellular wireless access networks,'' in \emph{Proc. IEEE
  Global Telecommunications Conf.}, Miami, FL, USA,  2010, pp. 1--5. DOI: 10.1109/GLOCOM.2010.5683654.

\bibitem{liu2015small}
C.~Liu, B.~Natarajan, and H.~Xia, ``Small cell base station sleep strategies
  for energy efficiency," \emph{IEEE Transactions on Vehicular Technology},
  vol.~65, no.~3, pp. 1652--1661, 2015.



%

\bibitem{Guo2016delay}
X. Guo, Z. Niu, S. Zhou, and P. R. Kumar, ``Delay-Constrained Energy-Optimal Base Station Sleeping Control," \emph{IEEE Journal on Selected Areas in Communications}, vol. 34, no. 5, pp. 1073--1085, 2016.


\bibitem{Sheng2014}
M. Sheng, C. Yang, Y. Zhang, and J. Li, ``Zone-Based Load Balancing in LTE Self-Optimizing Networks: A Game-Theoretic Approach," \emph{IEEE Transactions on Vehicular Technology}, vol. 63, no. 6, pp. 2916--2925, 2014.


\bibitem{xu2019load}
Y. Xu, W. Xu, Z. Wang, J. Lin, and S. Cui, ``Load Balancing for Ultradense Networks: A Deep Reinforcement Learning-Based Approach," \emph{IEEE Internet of Things Journal}, vol. 6, no. 6, pp. 9399--9412, 2019.

\bibitem{Attiah2020}
K. Attiah, K Banawan, A Gaber,  \emph{et~al.},  ``Load Balancing in Cellular Networks: A Reinforcement Learning Approach,'' in
  \emph{IEEE 17th Annual Consumer Communications \& Networking Conference (CCNC)}, Las Vegas, NV, USA, 2020, pp. 1--6. DOI: 10.1109/CCNC46108.2020.9045699.


\bibitem{Alsuhli2021}
G. Alsuhli, H. A. Ismail, K. Alansary, M. Rumman,  \emph{et~al.},  ``Deep Reinforcement Learning-based CIO and Energy Control for LTE Mobility Load Balancing,'' in
  \emph{IEEE 18th Annual Consumer Communications \& Networking Conference (CCNC)}, Las Vegas, NV, USA, 2021, pp. 1--6. DOI: 10.1109/CCNC49032.2021.9369525.


\bibitem{Alsuhli2021opti}
G. Alsuhli, K. Banawan, K. Seddik, and A. Elezabi,  ``Optimized Power and Cell Individual Offset for Cellular Load Balancing via Reinforcement Learning,'' in
  \emph{IEEE Wireless Communications and Networking Conference (WCNC)}, Nanjing, China, 2021, pp. 1--7. DOI: 10.1109/WCNC49053.2021.9417360.


\bibitem{Hasan2018}
M. M. Hasan, S. Kwon, and J. H. Na, ``Adaptive Mobility Load Balancing Algorithm for LTE Small-Cell Networks," \emph{IEEE Transactions on Wireless Communications}, vol. 17, no. 4, pp. 2205--2217, 2018.


\bibitem{Alsuhli2023}
G. Alsuhli, K. Banawan, A. Elezabi,  \emph{et~al.},``Mobility Load Management in Cellular Networks: A Deep Reinforcement Learning Approach," \emph{IEEE Transactions on Mobile Computing}, vol. 22, no. 3, pp. 1581--1598, 2023.



\bibitem{Interpretability1}
S. Chakraborty,  \emph{et~al.},``Interpretability of deep learning models: A survey of results," in \emph{2017 IEEE SmartWorld/SCALCOM/UIC/ATC/CBDCom/IOP/SCI}, San Francisco, CA, USA, 2017, pp. 1--6. DOI: 10.1109/UIC-ATC.2017.8397411.

\bibitem{Interpretability2}
R. Guidotti, A. Monreale, S. Ruggieri,  \emph{et~al.},``A survey of methods for explaining black box models," \emph{ACM computing surveys (CSUR)}, vol. 51, no. 5, pp. 1--42, 2018.

\bibitem{scalability1}
A.B. Bondi,``Characteristics of scalability and their impact on performance," in \emph{Proceedings of the 2nd international workshop on Software and performance}, 2000, pp. 195--203.


\bibitem{dawoud2014optimizing}
S.~Dawoud, A.~Uzun, S.~G{\"o}nd{\"o}r, and A.~K{\"u}pper, ``Optimizing the
  power consumption of mobile networks based on traffic prediction,'' in
  \emph{Proc. 38th IEEE  Annual Computer Software and Applications
  Conf.}, Vasteras, Sweden, 2014, pp. 279--288. DOI: 10.1109/COMPSAC.2014.38.

\bibitem{wang2017spatiotemporal}
J.~Wang, J.~Tang, Z.~Xu, Y.~Wang,  \emph{et~al.},
  ``Spatiotemporal modeling and prediction in cellular networks: A big data
  enabled deep learning approach,'' in \emph{Proc. IEEE Conf.
  on Computer Communications}, Atlanta, GA, USA, 2017, pp. 1--9. DOI: 10.1109/INFOCOM.2017.8057090.

\bibitem{wang2018spatio}
X.~Wang, Z.~Zhou, F.~Xiao,  \emph{et~al.},
  ``Spatio-temporal analysis and prediction of cellular traffic in
  metropolis,'' \emph{IEEE Transactions on Mobile Computing}, vol.~18, no.~9,
  pp. 2190--2202, 2018.

\bibitem{Xu2019wireless}
 Y. Xu, F. Yin, W. Xu, J. Lin, and S. Cui,
  ``Wireless Traffic Prediction With Scalable Gaussian Process: Framework, Algorithms, and Verification,'' \emph{IEEE Journal on Selected Areas in Communications}, vol.~37, no.~6,
  pp. 1291--1306, 2019.


\bibitem{GC-ZL-LG:14}
G.~Chen, Z.~Liu, and L.~Guo, ``The smallest possible interaction radius for
  synchronization of self-propelled particles,'' \emph{SIAM Review}, vol.~56,
  no.~3, pp. 499--521, 2014.

\bibitem{GC:17b}
G.~Chen, ``Small noise may diversify collective motion in vicsek model,''
  \emph{IEEE Transactions on Automatic Control}, vol.~62, no.~2, pp. 636--651,
  2017.

\bibitem{GL-CG:17}
G.~Chen, C.~Chen, G.~Yin, and L.~Y. Wang, ``Critical connectivity and fastest
  convergence rates of distributed consensus with switching topologies and
  additive noises,'' \emph{IEEE Transactions on Automatic Control}, vol.~62,
  no.~12, pp. 6152--6167, 2017.

\bibitem{8447261}
G.~Chen, X.~Duan, W.~Mei, and F.~Bullo, ``Linear stochastic approximation
  algorithms and group consensus over random signed networks,'' \emph{IEEE
  Transactions on Automatic Control}, vol.~64, no.~5, pp. 1874--1889, 2019.

\bibitem{CYBENKO1989279}
G.~Cybenko, ``Dynamic load balancing for distributed memory multiprocessors,''
  \emph{Journal of Parallel and Distributed Computing}, vol.~7, no.~2, pp.
  279--301, 1989. [Online]. Available: https://www.sciencedirect.com/science/article/pii/074373158990021X


\bibitem{sarkar2003survey}
T.~K. Sarkar, Z.~Ji, K.~Kim, A.~Medouri, and M.~Salazar-Palma, ``A survey of
  various propagation models for mobile communication,'' \emph{IEEE Antennas
  and propagation Magazine}, vol.~45, no.~3, pp. 51--82, 2003.

\bibitem{hamim2014overview}
S.~Hamim and M.~F. Jamlos, ``An overview of outdoor propagation prediction
  models,'' in \emph{Proc. Int. Symp. on Telecommunication
  Technologies}, Langkawi, Malaysia, 2014, pp.385--388. DOI: 10.1109/ISTT.2014.7238240.



 \bibitem{Chou2014}
S. F. Chou, T. C. Chiu, Y. J. Yu, and A. C. Pang, ``Mobile small cell deployment for next generation cellular networks,'' in \emph{2014 IEEE Global Communications Conference}, Austin, TX, USA, 2014, pp.4852--4857. DOI: 10.1109/GLOCOM.2014.7037574.


\bibitem{Smith1997}
K. Smith and M. Palaniswami,  ``Static and dynamic channel assignment using neural networks,''
  \emph{IEEE Journal on Selected Areas in Communications}, vol.~15, no.~2, pp. 238--249, 1997.


\bibitem{weingarten2004capacity}
H.~Weingarten, Y.~Steinberg, and S.~Shamai, ``The capacity region of the
  gaussian mimo broadcast channel,'' in \emph{Proc. Int. Symp.
  on Information Theory}, Chicago, IL, USA,  2004, pp. 174. DOI: 10.1109/ISIT.2004.1365211.

\bibitem{siomina2006automated}
I.~Siomina, P.~Varbrand, and D.~Yuan, ``Automated optimization of service
  coverage and base station antenna configuration in umts networks,''
  \emph{IEEE Wireless Communications}, vol.~13, no.~6, pp. 16--25, 2006.

%


\bibitem{ahuja2014network}
K.~Ahuja, B.~Singh, and R.~Khanna, ``Network selection algorithm based on link
  quality parameters for heterogeneous wireless networks,'' \emph{Optik}, vol.
  125, no.~14, pp. 3657--3662, 2014.




\bibitem{jangsher2015joint}
S.~Jangsher, H.~Zhou, V.~O. Li, and K.-C. Leung, ``Joint allocation of resource
  blocks, power, and energy-harvesting relays in cellular networks,''
  \emph{IEEE Journal on Selected Areas in Communications}, vol.~33, no.~3, pp.
  482--495, 2015.

\bibitem{semprebom2015sleep}
T.~Semprebom, C.~Montez, G.~Ara{\'u}jo, and P.~Portugal, ``A sleep-scheduling
  scheme for enhancing QoS and network coverage in IEEE 802.15. 4 WSN,'' in
  \emph{Proc. IEEE World Conf. on Factory Communication Systems}, Palma de Mallorca, Spain, 2015, pp. 1--4. DOI: 10.1109/WFCS.2015.7160579.

\bibitem{kashi2012coverage}
S.~S. Kashi and M.~Sharifi, ``Coverage rate calculation in wireless sensor
  networks,'' \emph{Computing}, vol.~94, no.~11, pp. 833--856, 2012.

\bibitem{shan2021intelligent}
C.~G. Shan, Z.~J. Ming, H.~X. Yi, L.~Nan, and L.~Lei, ``Intelligent mobile
  communication network plan for 5G based on insight big data,'' in
  \emph{Proc.  7th Int. Conf. on Signal and Information Processing, Networking and Computers}, Springer Singapore, 2021, pp. 1068--1074.

\bibitem{wang2012on}
G. Wang, H. Chen, Y. Li and M. Jin, ``On Received-Signal-Strength Based Localization with Unknown Transmit Power and Path Loss Exponent,'' \emph{IEEE Wireless Communications Letters}, vol.~1, no.~5, pp.
  536--539, 2012.

\bibitem{ts1996}
T. S. Rappaport, {\it{Wireless Communications: Principles and Practice}}, Prentice-Hall, 1996.

\bibitem{strang2006linear}
G.~Strang, {\it{Linear algebra and its applications}}, 4th ed.  Belmont, CA: Thomson, Brooks/Cole, 2006.

\bibitem{bullo2020lectures}
F.~Bullo, {\it{Lectures on network systems}}, 1.4 ed. Kindle Direct Publishing, 2019.

\bibitem{lee2013stochastic}
C.-H. Lee, C.-Y. Shih, and Y.-S. Chen, ``Stochastic geometry based models for
  modeling cellular networks in urban areas,'' \emph{Wireless networks},
  vol.~19, no.~6, pp. 1063--1072, 2013.

\bibitem{guo2013spatial}
A.~Guo and M.~Haenggi, ``Spatial stochastic models and metrics for the
  structure of base stations in cellular networks,'' \emph{IEEE Transactions on
  Wireless Communications}, vol.~12, no.~11, pp. 5800--5812, 2013.

\bibitem{riihijarvi2010modeling}
J.~Riihijarvi and P.~Mahonen, ``Modeling spatial structure of wireless
  communication networks,'' in \emph{Proc. IEEE Conf. on Computer
  Communications Workshops}, San Diego, CA, USA, 2010, pp. 1--6. DOI: 10.1109/INFCOMW.2010.5466702.

\bibitem{zorich2016mathematical}
V.~A. Zorich,  {\it{Mathematical analysis I}}, 2nd ed. Berlin: Springer, 2016.

\bibitem{Pinkus1999}
A. Pinkus, ``Approximation theory of the MLP model in neural networks,'' \emph{Acta Numerica}, vol. 9, pp. 143--195, 1999.


\bibitem{Attali1997}
J. G.~Attali, G.~Pagès, ``Approximations of functions by a multilayer perceptron: a new approach,'' \emph{Neural networks}, vol. 10, no.~6, pp. 1069--1081, 1997.


\bibitem{ruby2020binary}
U.~Ruby and V.~Yendapalli, ``Binary cross entropy with deep learning technique
  for image classification,'' \emph{Int. J. Adv. Trends Comput. Sci. Eng},
  vol.~9, no.~10, pp. 5395--5397, 2020.

\bibitem{Leland1994}
W. Leland, M. Taqqu, W. Willinger, and D. Wilson, ``On the self-similar nature of Ethernet traffic (extended version)," \emph{IEEE/ACM Transactions on networking}, vol. 2, no. 1, pp. 1--15, 1994.

\bibitem{touchette2009large}
H. Touchette, ``The large deviation approach to statistical mechanics," \emph{Physics Reports}, vol. 478, no. 1, pp. 1--69, 2009.]


\bibitem{zhao2018deployment}
H. Zhao, H. Wang, W. Wu, and J. Wei, ``Deployment Algorithms for UAV Airborne Networks Toward On-Demand Coverage," \emph{IEEE Journal on Selected Areas in Communications}, vol. 36, no. 9, pp. 2015--2031, 2018.


\end{thebibliography}
\end{document}